\documentclass[aps, prd, twocolumn, 
showpacs, showkeys, 
print, preprintnumbers, nofootinbib,
groupedaddress
]{revtex4-2}
\usepackage[utf8]{inputenc}
\usepackage[T1]{fontenc}
\usepackage{lmodern}
\usepackage{microtype}
\usepackage[english]{babel}

\usepackage{amssymb,amsbsy,amsmath,amsfonts,amsthm}
\usepackage{mathtools}
\allowdisplaybreaks
\usepackage{yhmath} 

\usepackage{slashed}
\usepackage{bm}
\usepackage{times}
\usepackage{soul}
\setstcolor{red}

\usepackage[dvipsnames,table,xcdraw]{xcolor}

\usepackage{graphicx}
\usepackage{epstopdf}
\usepackage{tabularx}
\usepackage{multirow}
\usepackage{longtable}

\usepackage{hyperref}
\hypersetup{ colorlinks=true, linkcolor=blue, citecolor=cyan, urlcolor=red
}

\usepackage{url}
\usepackage{orcidlink}

\usepackage{comment}

\begin{document}

\title{Heavy $K^*$ mesons with open charm from $KD^{(*)}D^*$ interactions}

\author{Xiu-Lei Ren\orcidlink{0000-0002-5138-7415}}
\email[]{xiulei.ren@uni-mainz.de}
\affiliation{Helmholtz Institut Mainz,   D-55099 Mainz, Germany }

\author{K. P. Khemchandani\orcidlink{0000-0003-2686-5419}}
\email[]{kanchan.khemchandani@unifesp.br}
\affiliation{Universidade Federal de S\~ao Paulo, C.P. 01302-907, S\~ao Paulo, Brazil}

\author{A. Mart\'inez Torres\orcidlink{0000-0002-3656-5834}}
\email[]{amartine@if.usp.br}
\affiliation{Instituto de F\'isica, Universidade de S\~ao Paulo, C.P. 66318, 05389-970 S\~ao 
Paulo, S\~ao Paulo, Brazil}

\begin{abstract}
In view of the recent discovery of $T_{cc}$, which can be described as a $DD^*$ molecular state, we perform a study of the $KDD^*$ system, and its extension to studying $KD^*D^*$, where $D^*D^*$ is bound as a spin 1 $T_{cc}$-like state, to search for the possible existence of exotic mesons with the open flavors $ccs$ being part of their quark configuration. Considering the additional attractive interactions present in the $KD$ and $KD^*$ subsystems, where the states $D^*_{s0}(2317)$ and $D_{s1}(2460)$ are generated, we solve the Faddeev equations considering the fixed-center approximation for the mentioned three-body systems and find the existence of two doubly charmed $K^*$-like mesons, $K_{cc}^*(4309)$ and $K_{cc}^*(4449)$, with quantum numbers $I(J^P)=1/2\,(1^-)$. Considering the respective three-body thresholds of the two systems, both $K_{cc}$-states are bound by around $60$ MeV. An experimental confirmation will bring evidence for the existence of a degree of exoticity beyond charm +2.
\end{abstract}


\maketitle

\date{\today}

\section{Introduction}

The search for the existence of exotic hadrons is at the forefront of hadron physics, deepening our understanding of quantum chromodynamics (QCD). A major breakthrough occurred in 2003 when the charmonium-like meson $\chi_{c1}(3872)$ was discovered by the Belle Collaboration~\cite{Belle:2003nnu}. After one decade, the charged charmonium meson $Z_c(3900)$, the first confirmed hadron with a minimal four-quark content, was observed by the BESIII and Belle experiments~\cite{BESIII:2013ris,Belle:2013yex}. Most recently, the LHCb Collaboration reported the observation of the $T_{cc}^+$ state~\cite{LHCb:2021auc}, which is the first doubly charmed exotic meson observed. Along with a series of XYZ states, other resonances such as the lightest positive parity charmed states $D_{s0}^*(2317)$ and $D_{s1}(2460)$~\cite{BaBar:2003oey,CLEO:2003ggt}, the fully charmed tetraquark candidate $X(6900)$~\cite{LHCb:2020bwg} and the $P_c$ pentaquarks~\cite{LHCb:2015yax,LHCb:2019kea}, the existence of exotic hadrons has expanded our knowledge of the working of strong interactions between heavy quarks, as well as between heavy and light quarks,but particularly in the charm sector~\cite{Olsen:2017bmm,Brambilla:2019esw}. 

From the theoretical side, there are tremendous efforts in the hadron physics community to explain the nature of exotic charm hadrons (we refer the interested readers to the most recent reviews for more details~\cite{Meng:2022ozq,Mai:2022eur,Chen:2022asf,Brambilla:2019esw,Liu:2019zoy,Guo:2017jvc}). Here we would like to emphasize the relevance of the hadronic molecule picture, which provides a natural explanation of the properties observed for many of the exotic states found as two-hadron molecules. For instance, $\chi_{c1}(3872)$ and $Z_c(3900)$ are expected to be $s$-wave $D\bar{D}^*+c.c.$ molecules, and $T_{cc}^+$ can be interpreted as a bound state of the $DD^*$ system~\cite{Gamermann:2007fi,Guo:2013zbw,Aceti:2014uea,Gamermann:2009uq,Ling:2021bir,Du:2022jjv,Dai:2023kwv}. 
As a natural extension of the existence of hadron molecules of a two-body nature, one can expect the existence of three-body molecules or multi-body states in the charm sector, likewise the deuteron, triton, and nucleus in nuclear physics. Such kind of research tendency has been highlighted in the last years, for instance, in Refs.~\cite{MartinezTorres:2020hus,Liu:2024uxn}.

Recently, the interest in the existence of exotic mesons with strangeness seems to grow increasingly, and several theoretical studies have been made investigating the interactions between mesons with strangeness, like $K$, and charm mesons like $D$/$\bar{D}$, $D^*$/$\bar{D}^*$~\cite{Ma:2017ery,Ren:2018pcd,Wu:2020job,Wei:2022jgc,SanchezSanchez:2017xtl,MartinezTorres:2018zbl,Wu:2019vsy,Pang:2020pkl,Xiao:2024dyw,Belle:2020xca,Tan:2024omp,Zhang:2024yfj}. These investigations are motivated by the strong s-wave attractive interactions present in the $KD$ and $KD^*$ systems, which give rise to the $D_{s0}^*(2317)$ and $D_{s1}(2460)$ resonances, respectively~\cite{Guo:2006fu,Gamermann:2006nm}. 
Having strange and charm quarks, different super-exotic structures, non-compatible with the quantum numbers obtained from the traditional quark model, can emerge in these systems, like heavy $K^*$ hexaquarks with hidden charm ($c\bar{c}s\bar{q}q\bar{q}$) or doubly charmed states with strangeness ($ccs\bar{q}\bar{q}\bar{q}$).

The $KD\bar{D}^*$ and $KD\bar{D}$ systems belong to the former mentioned case, i.e., hidden charm and open strangeness. As shown in  Refs.~\cite{Ma:2017ery,Ren:2018pcd,Wu:2020job,Wei:2022jgc}, considering s-wave interactions, two hidden charm $K_{c\bar{c}}^*$ mesons are found. In particular, a state about 50-80 MeV below the three-body threshold is obtained in the $KD\bar{D}^*$ system with quantum numbers of isospin $I$ and spin-parity $J^P$ given by $I(J^P)=1/2(1^-)$. In the case of the $KD\bar{D}$ system, a state with binding energy around $50$ MeV and  $I(J^P)=1/2(0^-)$ is found as a consequence of the three-body dynamics involved in these systems.  
In the case of the even more exotic quantum numbers of open strangeness and double charm, the existence of such states has been claimed in Refs.~\cite{SanchezSanchez:2017xtl,MartinezTorres:2018zbl,Wu:2019vsy} from the dynamics involved in the $KDD$ system by using different approaches~\cite{SanchezSanchez:2017xtl,MartinezTorres:2018zbl,Wu:2019vsy}. 
In both the studies, a doubly charmed $K_{cc}^*$ bound state is predicted with a binding energy around $65-90$ MeV~\cite{MartinezTorres:2018zbl,Wu:2019vsy}. Such a system has also been studied in a finite volume~\cite{Pang:2020pkl,Xiao:2024dyw} and experimental searches for such a state have already started~\cite{Belle:2020xca}, although better statistics are necessary to reach to any conclusion. The $KDD^*$ has been investigated in Ref.~\cite{Ma:2017ery} by solving the Schr\"{o}dinger equation with the Born-Oppenheimer approximation, and, after the discovery of $T_{cc}^+$, the interest in such a system has grown. Most recently, the former system has been investigated by using the chiral quark model~\cite{Tan:2024omp} and the  lattice effective field theory~\cite{Zhang:2024yfj}. The result of Ref.~\cite{Zhang:2024yfj} confirmed the existence of a bound state as claimed in Ref.~\cite{Ma:2017ery}, with a binding energy around $44\sim 84$ MeV. Furthermore, the $KD^*D^*$ system has been investigated in Ref.~\cite{SanchezSanchez:2017xtl} by using the $D_{s1}D^*$ quasi-two body scattering, finding loosely bound states with spin-parity $J^P=0^-,2^-$. Similarly, other three-body systems, such as those involving the vector meson $K^*$ instead of the pseudoscalar $K$ meson, and/or their extension to the bottom sector, with the $B/\bar{B}$, $B^*/\bar{B}^*$ pairs instead of the $D$/$\bar D$, $D^*/\bar D^*$ mesons have been investigated in Refs.~\cite{Valderrama:2018knt,Ren:2018qhr,Ikeno:2022jbb,Bayar:2023itf}. 

It is in order here to mention that despite having experimental access to energies $\simeq$ 4000 MeV, finding of no new $K$ or $K^*$ mesons have been claimed. A quick look at the review of the Particle Data Book (PDG), shows an inactivity in the spectroscopy of strange mesons, with $K^*(1680)$ being the latest $J^P=1^-$ state with strangeness observed. Observation of some of the aforementioned exotic mesons with strangeness has been claimed in the $J/\psi\pi K$ invariant mass distribution obtained from weak decays of the $B$ meson~\cite{Ren:2019rts}, and it is simply a matter of time before experimental investigations of such mesons will be carried out.

Inspired by the growing interest in the spectroscopy of exotic mesons with strangeness, we extend our previous investigation of the $KD\bar{D}^*$~\cite{Ren:2018pcd} system to analyze the formation of three-body states in the $KDD^*$ system. This is done by using the fixed-center approximation (FCA) to solve the Faddeev equations~\cite{Deloff:1999gc,Kamalov:2000iy}, an approximation which is valid for studying the formation of states below the three-body threshold of a system where its dynamics can be considered as that of a particle interacting with a heavier cluster~(for detailed discussions, see the review in Ref.~\cite{MartinezTorres:2020hus} and for applicability limits see Refs.~\cite{MartinezTorres:2010ax,Malabarba:2021taj}). Having this in mind, we consider in this case that the heavy $DD^*$ system clusters as $T_{cc}$~\cite{Feijoo:2021ppq}, and the light $K$-meson scatters off the $D$ and $D^*$ mesons forming $T_{cc}$. Due to the description of  $D_{s0}^*(2317)$, $D_{s1}(2460)$ and $T_{cc}$  as states generated from the $KD$, $KD^*$ and $DD^*$ (coupled-channel) interactions, respectively, a doubly charmed bound state with positive strangeness is obtained. Besides, we consider that the generation of a $T_{cc}$-like state is also predicted from the interaction of $D^*D^*$ with $J^P=1^+$ in Ref.~\cite{Dai:2021vgf} by solving the Bethe-Salpeter equation with a coupled channels approach. Thus, we find it interesting to extend our FCA study to the $KD^*D^*$ system and explore the possible existence of a bound state. 

The paper is organized as follows. In Sec.~\ref{Sec:FCA}, we outline the main aspects of the formalism employed to investigate the three-body systems. The results obtained for the $KDD^*$ and $KD^*D^*$ systems, along with the corresponding discussion, are presented in Sec.~\ref{Sec:Results}. Finally, we summarize the main conclusions obtained in Sec.~\ref{Sec:Summary}.

\section{Fixed-center approximation to Faddeev equations}\label{Sec:FCA}

Generally, three-hadron systems can be studied by solving the three-coupled integral Faddeev equations~\cite{Faddeev:1960su}, which allows to determine the $T$-matrix of the system as a sum of three partitions $T_1$, $T_2$ and $T_3$, each of them containing the information of the scattering where particle $i=1,2,3$ starts being a spectator in the lowest order contribution to the scattering series in $T_i$. However, solving such equations is not an easy task, especially, when coupled channels are involved, and, in many occasions, approximate methods are often implemented to simplify the Faddeev equations. In this context, the relevance of the use of effective Lagrangians to determine the input two-body $t$-matrices by considering the on-shell factorization of the Bethe-Salpeter equation and the use of these $t$-matrices as input for the Faddeev equations, have allowed studying a large variety of three-body systems with a large number of coupled channels~\cite{MartinezTorres:2007sr,MartinezTorres:2008gy,Khemchandani:2008rk}. 

\begin{figure}[t]
  \includegraphics[width=0.49\textwidth]{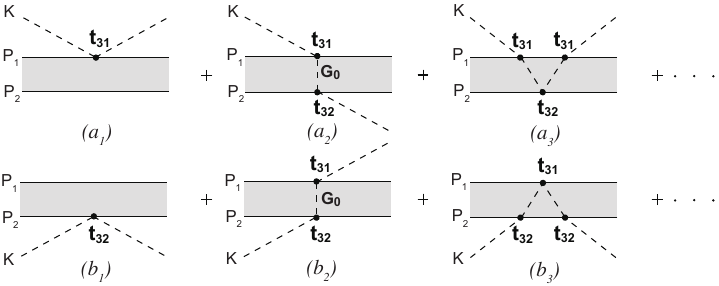}
  \caption{Diagrams representing the $K$ meson scattering off the $R_{12}$ cluster constituted by the $P_1$ and $P_2$ particles. The rescattering diagrams in the first row provide the Faddeev partition $T_{31}$, and the second-row diagrams give rise to $T_{32}$.}
  \label{Fig:FCA}
\end{figure}

Within the different approximate methods to solve the Faddeev equations, there is one which is especially suitable for studying the scattering between a cluster ($R_{12}$) composed of two particles, which we denote as particles $P_1$ and $P_2$, whose mass is larger than the other particle of the system, which we denote as particle $P_3$ (e.g., the $K$ meson in the current work), which is known as the fixed center approximation~\cite{Chand:1962ec,Barrett:1999cw,Deloff:1999gc,Kamalov:2000iy}. In this case, the scattering between the three particles resembles that of a particle with a fixed, heavier, scattering center formed by the other two particles and such a description is a rather good approximation when studying the formation of three-body states below the three-body threshold~\cite{Toker:1981zh,Gal:2006cw}. For the technical details of FCA, we refer the reader to Refs.~\cite{Roca:2010tf,Yamagata-Sekihara:2010muv,MartinezTorres:2020hus}. Here, we summarize the main aspects of the formalism. Within the fixed-center approximation, the scattering consists of a series of contributions where particle 3 scatters off particles 1 and 2 forming the cluster, having then two series of contributions (see Fig.~\ref{Fig:FCA}): those where particle 3 interacts first with particle 1 of the cluster and starts rescattering with the two particles forming the cluster ($T_{31}$ partition); Those contributions in which particle 3 starts interacting first with particle 2 of the cluster and keeps rescattering with the particles of the cluster ($T_{32}$ partition). Then, the total scattering amplitude $T_\text{tol}$ is given as follows: 
\begin{equation}\label{Eq:FCA}
\begin{aligned}
	T_\mathrm{tol} &= T_{31} + T_{32}, \\
	T_{31} &= t_{31} +  t_{31}\, G_0\, t_{32} +  t_{31}\, G_0\, t_{32}\, G_0\, t_{31} + \cdots\\ 
	& \equiv  t_{31} + t_{31}\, G_0\, T_{32} ,\\
	T_{32} &= t_{32} + t_{32}\, G_0\, t_{31} + t_{32}\, G_0\, t_{31} \, G_0\,  t_{32} + \cdots \\ 
	&\equiv t_{32} + t_{32}\, G_0\, T_{31}.
\end{aligned}
\end{equation}
As illustrated in Figs.~\ref{Fig:FCA}-($a_1$) and -($b_1$), the amplitudes $t_{31}$ and $t_{32}$ represent the single scattering of the $K$ meson with the $R_{12}$ cluster: $t_{31}$ denoting the $K$-$P_1$ interaction and $P_2$ as a spectator, and $t_{32}$ for the $K$-$P_2$ interaction with $P_1$ as a spectator. 
 Through the rescattering procedure, $K$ can propagate through the cluster to trigger the double scattering ($ t_{31}\, G_0\, t_{32}$ and $ t_{32}\, G_0\, t_{31} $), the triple scattering, etc. The loop function $G_0$ in Eq.~\eqref{Eq:FCA} represents the Green function of the $K$ meson propagating in the cluster $R_{12}$. After solving the coupled equations in~Eq.~\eqref{Eq:FCA}, which are algebraic equations, the total three-body amplitude can be written as:
\begin{equation}\label{Eq:Ttol}
	T_\mathrm{tol} = \frac{t_{31}+t_{32} + 2 t_{31}t_{32} G_0}{1-t_{31}t_{32}G_0^2}.
\end{equation} 
In Eq.~(\ref{Eq:Ttol}), $t_{31}$ and $t_{32}$ depend, respectively, on the invariant masses of the (31) and (32) subsystems, which are functions of the center-of-mass energy, $\sqrt{s}$, and the mass of the cluster $R_{12}$, while $G_0$ depends on $\sqrt{s}$ and the mass of the cluster $R_{12}$. The latter is fixed to the mass of the particle obtained from the interaction of particles 2 and 3 (in this case, $T_{cc}$ or $T_{cc}$ like states). The $t_{31}$ and $t_{32}$ amplitudes, at the same time, depend on the weighted combinations of the different isospin $t$-matrices describing the (31) and (32) subsystems, respectively.

In the following, we present the steps followed to find the expression of $t_{31/32}$ in terms of the isospin $t$-matrices related to the (31)/(32) subsystems and provide the expression of $G_0$ for the specific three-body systems considered in the present work, i.e. $KDD^*$ and  $KD^*D^*$.

\subsection{Singe scattering amplitudes}
First, to illustrate the method, we focus on the $KDD^*$ system, where $DD^*$ is considered to form the $T_{cc}$ state with isospin $I=0$ and $J^P=1^+$~\cite{Feijoo:2021ppq}. In the isospin basis,  $T_{cc}$ is described by the ket $|DD^*;I=0,I_z=0\rangle$, which, in terms of the charged states, can be written as 
\begin{equation}
	|DD^*;I=0,I_z=0\rangle = -\frac{1}{\sqrt{2}} \bigl(|D^+D^{*0}\rangle - |D^0 D^{*+}\rangle \bigr),
\end{equation}  
with the isospin doublets being: $(D^+,-D^0)$ and $(D^{*+}, -D^{*0})$. If we add now a Kaon, we have the following isospin ket for the three-body $KDD^*$ system: 
\begin{equation}
	\left| K(DD^*); I=\frac{1}{2}, I_z=\frac{1}{2}\right\rangle = \biggl|K; \frac{1}{2},\frac{1}{2}\biggr\rangle \otimes \biggl|DD^*; 0, 0 \biggr \rangle,
\end{equation}
where the total isospin of $KDD^*$ is $I=1/2$ and the third component of isospin is chosen, for example, as $I_z=1/2$.  
The matrix elements of the single scattering amplitudes $t_{31}$ and $t_{32}$ can be evaluated via  
\begin{equation}\label{Eq:cal_t31t32}
\begin{aligned}
	t_{31} & = \langle \underbracket[0.5pt][2.pt]{K(D}D^*); 1/2, 1/2 | \hat{t}_{31} |\underbracket[0.5pt][2pt]{K(D} D^*); 1/2, 1/2\rangle, \\
	t_{32} &= \langle  \underbracket[0.5pt][2pt]{K(DD^*}); 1/2 , 1/2 | \hat{t}_{32} | \underbracket[0.5pt][2pt]{K(DD^*}); 1/2, 1/2\rangle,
\end{aligned}
\end{equation}
where the under brackets in the ket denote the collision configurations: $KD$ (i.e., particles 3 and 1, respectively) and $KD^*$ (particles 3 and 2). Next, we express the kets in Eq.~\eqref{Eq:cal_t31t32} in terms of isospin kets related to the subsystems (31) and (32), depending on whether we evaluate $t_{31}$ or $t_{32}$: 
\begin{equation}
\begin{aligned}
	& |\underbracket[0.5pt][2pt]{K(D} D^*); 1/2, 1/2\rangle \\
	&= \frac{1}{\sqrt{2}} \Biggl[ \left|I^{KD}=1, I_z^{KD}=1\right\rangle \otimes \left|I_z^{D^*} = -\frac{1}{2}\right\rangle \\
	&\quad  - \frac{1}{\sqrt{2}} \biggl( \left|I^{KD}=1,I_z^{KD}=0\right\rangle + \left|I^{KD}=0,I_z^{KD}=0\right\rangle \biggr) \\
	&\qquad \otimes\left|I_z^{D^*}=\frac{1}{2} \right\rangle 
	\Biggr]\, ,
\end{aligned}
\end{equation}	
\begin{equation}
\begin{aligned}
	& |\underbracket[0.5pt][2pt]{K(D D^*}); 1/2, 1/2\rangle \\
	&= - \frac{1}{\sqrt{2}} \Biggl[  \left|I^{KD^*}=1, I_z^{KD^*}=1\right\rangle \otimes \left|I_z^{D} = -\frac{1}{2}\right\rangle \\
	&\quad  - \frac{1}{\sqrt{2}} \biggl( \left|I^{KD^*}=1,I_z^{KD^*}=0\right\rangle + \left|I^{KD^*}=0,I_z^{KD^*}=0\right\rangle \biggr) \\
	&\qquad \otimes\left|I_z^{D}=\frac{1}{2} \right\rangle 
	\Biggr].
\end{aligned}
\end{equation}
By plugging the above kets into Eq.~\eqref{Eq:cal_t31t32}, we obtain the single scattering amplitudes $t_{31}$ and $t_{32}$ as combinations of the isospin $0$ and $1$ $s$-wave amplitudes of the $KD$ and $KD^*$ pairs:  
\begin{equation}\label{Eq:t3132inKDDx}
\begin{aligned}
	t_{31} & = \frac{t_{KD}^{I=0} + 3\, t_{KD}^{I=1}}{4}\, ,\\
	t_{32} & = \frac{t_{KD^*}^{I=0} + 3\, t_{KD^*}^{I=1}}{4}.
\end{aligned}
\end{equation}

Following the same procedure, one can easily obtain the single scattering amplitudes for the $K(D^*D^*)$ system, finding: 
\begin{equation}\label{Eq:t3132inKDxDx}
	t_{31}=t_{32} = \frac{t_{KD^*}^{I=0} + 3 t_{KD^*}^{I=1} }{4}.
\end{equation}

As can be seen in Eqs.~(\ref{Eq:t3132inKDDx}) and (\ref{Eq:t3132inKDxDx}), we need the two-body $t$-matrices describing the $KD$ and $KD^*$ interactions. These $t$-matrices can be determined by solving the coupled channel Bethe-Salpeter equations 
\begin{align}
t^I=v^I+v^Igt^I\label{BS}
\end{align}
for a certain isospin $I$ of the system. In Eq.~(\ref{BS}), $g$ is a diagonal matrix in the coupled-channel space whose elements $g_i$ are the two-hadron loop functions for the channels $i$. These $g_i$ are regularized by introducing subtraction constants and we consider the same values as those used in Refs.~\cite{Guo:2006fu,Guo:2006rp}. The kernel $v^I$ in Eq.~(\ref{BS}) is a matrix whose elements $v^{I}_{ij}$ describe the transition amplitude $i\to j$ and they are obtained from a Lagrangian based on the chiral and heavy quark symmetries. Here we follow the approach of Refs.~\cite{Guo:2006fu,Guo:2006rp}, where the $KD$ and $KD^*$ interaction kernels are determined from the lowest order chiral Lagrangian based on heavy-quark spin symmetry.

In the case of total isospin $I=0$, the $KD^{(*)}$ system, together with $\eta D_s^{(*)}$, are considered as couple channels when solving the Bethe-Salpeter equations. At leading order, $v^{I=0}$ is given by~\cite{Guo:2006fu,Guo:2006rp}:
\begin{equation}
  \begin{pmatrix}
      v_{KD,KD}^{I=0} & v_{KD,\eta D_s}^{I=0} \\
      v_{\eta D_s,KD}^{I=0} & v_{\eta D_s,\eta D_s}^{I=0} 
  \end{pmatrix} = 
  \begin{pmatrix}
      -2 & \sqrt{3} \\
      \sqrt{3} & 0
  \end{pmatrix} \frac{s-u}{4f_\pi^2},\label{V1}
\end{equation}
\begin{equation}
  \begin{pmatrix}
      v_{KD^*,KD^*}^{I=0} & V_{KD^*,\eta D_s^*}^{I=0} \\
      V_{\eta D_s^*,KD^*}^{I=0} & V_{\eta D_s,\eta D_s^*}^{I=0} 
  \end{pmatrix} = 
\begin{pmatrix}
      -2 & \sqrt{3} \\
      \sqrt{3} & 0
  \end{pmatrix} \frac{s-u}{4f_\pi^2}\varepsilon\cdot\varepsilon',\label{V2}
\end{equation}
where $f_\pi=92.4$ MeV is the pion decay constant, $s$ and $u$ are Mandelstam variables, and $\varepsilon^{(')}$ denotes the polarization vector associated with the vector meson in the initial (final) state. The amplitudes $v^I_{i,j}$ in Eqs.~(\ref{V1}) and (\ref{V2}) are further projected on the s-wave by considering 
\begin{align}
\frac{1}{2}\int\limits_{-1}^1\text{cos}\theta\, v^I_{i,j}(s,u(\theta)), 
\end{align}
with $\theta$ being the angle formed between the incident and the scattered Kaon in the center-of-mass frame. 

Similarly, for total isospin $I=1$, in Refs.~\cite{Guo:2006fu,Guo:2006rp}, $KD^{(*)}$ and $\pi D^{(*)}_s$ are treated as coupled channels when solving the Bethe-Salpeter equation, and 
at leading order, $v^{I=1}$ read as~\cite{Guo:2006fu,Guo:2006rp} 
\begin{equation}
  \begin{pmatrix}
      v_{KD,KD}^{I=1} & v_{KD,\pi D_s}^{I=1} \\
      v_{\pi D_s,KD}^{I=1} & v_{\pi D_s,\pi D_s}^{I=1} 
  \end{pmatrix} = 
  \begin{pmatrix}
      0 & -1 \\
      -1 & 0
  \end{pmatrix} \frac{s-u}{4f_\pi^2},
\end{equation}
\begin{equation}
  \begin{pmatrix}
      v_{KD^*,KD^*}^{I=1} & V_{KD^*,\pi D_s^*}^{I=1} \\
      V_{\pi D_s^*,KD^*}^{I=1} & V_{\pi D_s,\pi D_s^*}^{I=1} 
  \end{pmatrix} = 
\begin{pmatrix}
      0 & -1 \\
      -1 & 0
  \end{pmatrix} \frac{s-u}{4f_\pi^2}\varepsilon\cdot\varepsilon'.
\end{equation}

After projecting the aforementioned amplitudes in the $s$-wave, they are plug into Eq.~(\ref{BS}), obtaining in this way the two-body $t$-matrices $t_{KD}^{I=0,1}$ and $t_{KD^*}^{I=0,1}$ in Eqs.~\eqref{Eq:t3132inKDDx} and \eqref{Eq:t3132inKDxDx}. As a consequence of the dynamics considered, poles in the $t$-matrices for the $KD$ and $KD^*$ coupled-channel systems in the isospin 0 sector are found at 2312 MeV and $2452$ MeV, respectively, which can be associated with $D_{s0}^*(2317)$ and $D_{s1}(2460)$. The dominant molecular nature obtained for $D_{s0}^*(2317)$ and $D_{s1}(2460)$ has been confirmed from lattice QCD studies~\cite{MartinezTorres:2014kpc}. 

Furthermore, as a consequence of the different normalizations between the scattering matrix of a three-body system and that of a particle+cluster system, with the cluster arising from the interaction of two particles, we need to normalize the $t_{31}$ and $t_{32}$ amplitudes in Eq.~\eqref{Eq:FCA} before solving the equation. In particular, in this case~\cite{Ren:2018pcd}
\begin{equation}\label{Eq:WeightFactor}
	t_{31} \to  \frac{M_R}{m_1} t_{31},\quad t_{32} \to \frac{M_R}{m_2}t_{32},
\end{equation}
where $M_R$ represents the mass of the cluster, and $m_i$, $i=1,2$, the mass of the particle $P_i$ in the cluster. 

As a final comment in this section, we would like to mention that the arguments of $t_{31}$ and $t_{32}$ are $s_{31}$ and $s_{32}$, which are the invariant masses of the $KP_1$ and $KP_2$ systems. The latter can be related to the center-of-mass energy of the three-body system, $\sqrt{s}$, via the following equations~\cite{Khemchandani:2008rk}:   
\begin{equation}\label{Eq:s31s32}
\begin{aligned}
	s_{31} &= m_3^2 + m_1^2 + \frac{M_R^2+m_1^2-m_2^2}{2M_R^2} (s-m_3^2-M_R^2),\\
	s_{32} &= m_3^2 + m_2^2 + \frac{M_R^2+m_2^2-m_1^2}{2M_R^2} (s-m_3^2-M_R^2).
\end{aligned}	
\end{equation}

\begin{table}[t]
\centering
\caption{Values of the cluster mass $M_R$ and cutoff $\Lambda_{R}$ (in units of MeV) for the $K(DD^*)$ and $K(D^*D^*)$ systems.}
\label{Tab:MRLambR}
\begin{tabular} {cp{3.3cm}<{\centering}p{3.3cm}<{\centering}}
\hline\hline 
            & $K(DD^*)$ & $K(D^*D^*)$ \\ 
\hline
$M_R$      &  $3874.82$ &   $4015.54$  \\  
$\Lambda_R$ & $415$     &  $420$  \\  
\hline\hline 
\end{tabular}
\end{table}

\begin{figure}[t]
  \includegraphics[width=0.4\textwidth]{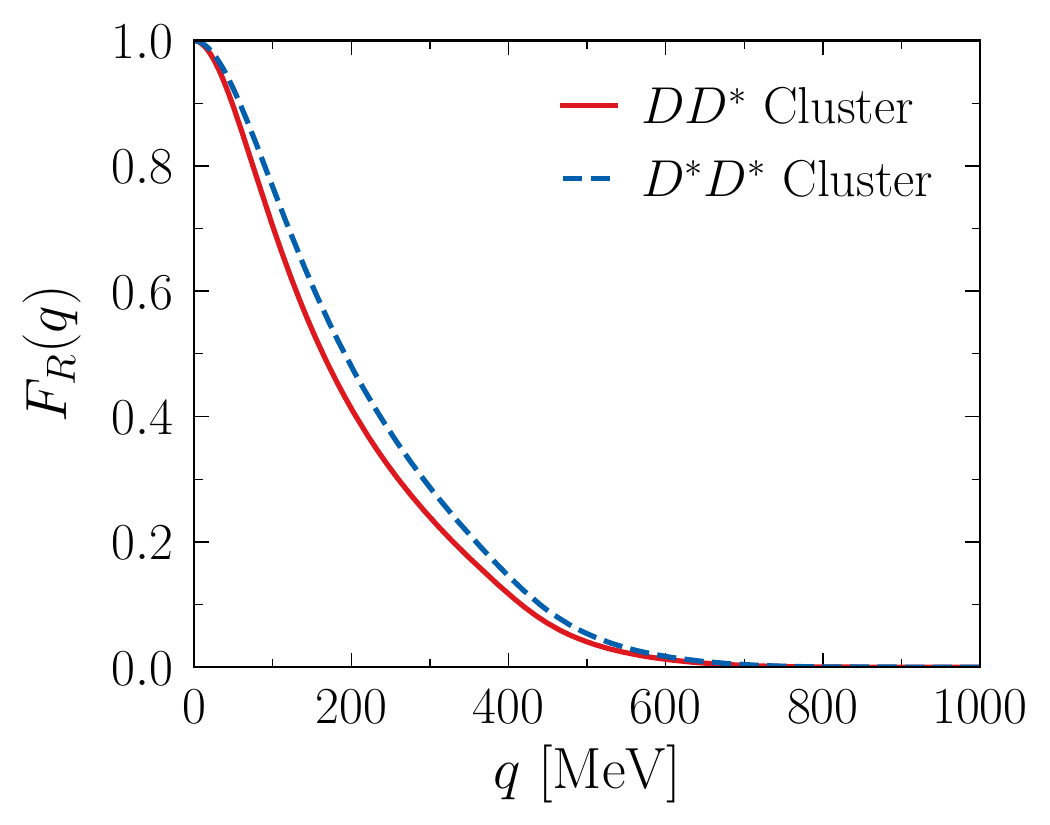}
  \caption{Form factors related to the $(DD^*)_{T_{cc}}$ cluster with $\Lambda_R=415$ MeV (solid line) and $(D^*D^*)_{T_{cc}}$ cluster with $\Lambda_R=420$ MeV (dashed line).}
  \label{Fig:FormFactor}
\end{figure}

\begin{figure}[t]
  \includegraphics[width=0.5\textwidth]{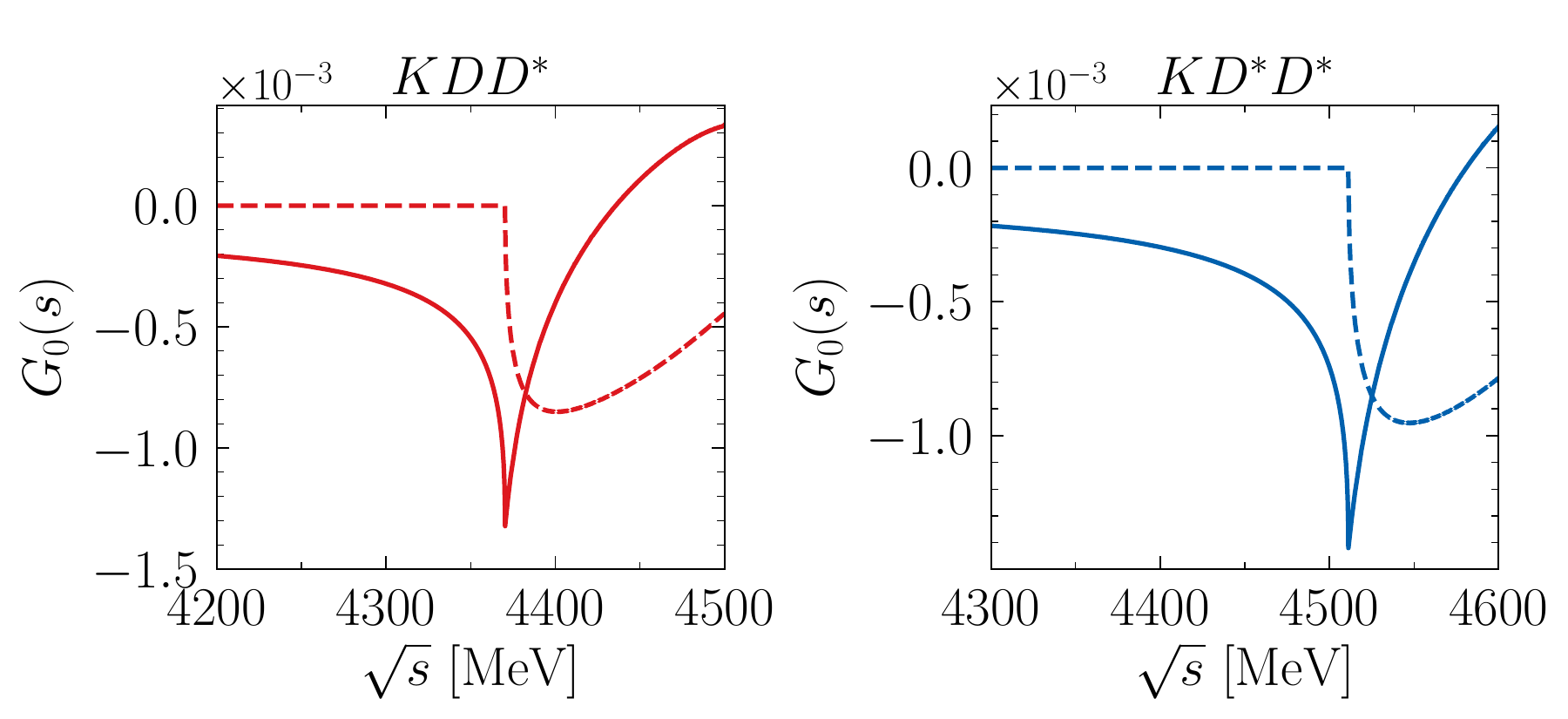}
  \caption{Real (solid line) and imaginary (dashed line) parts of $G_0(s)$ function in the $K(DD^*)$ system (left panel) and the $K(D^*D^*)$ system (right panel).}
  \label{Fig:G0}
\end{figure}

\begin{figure*}[t]
  \includegraphics[width=0.7\textwidth]{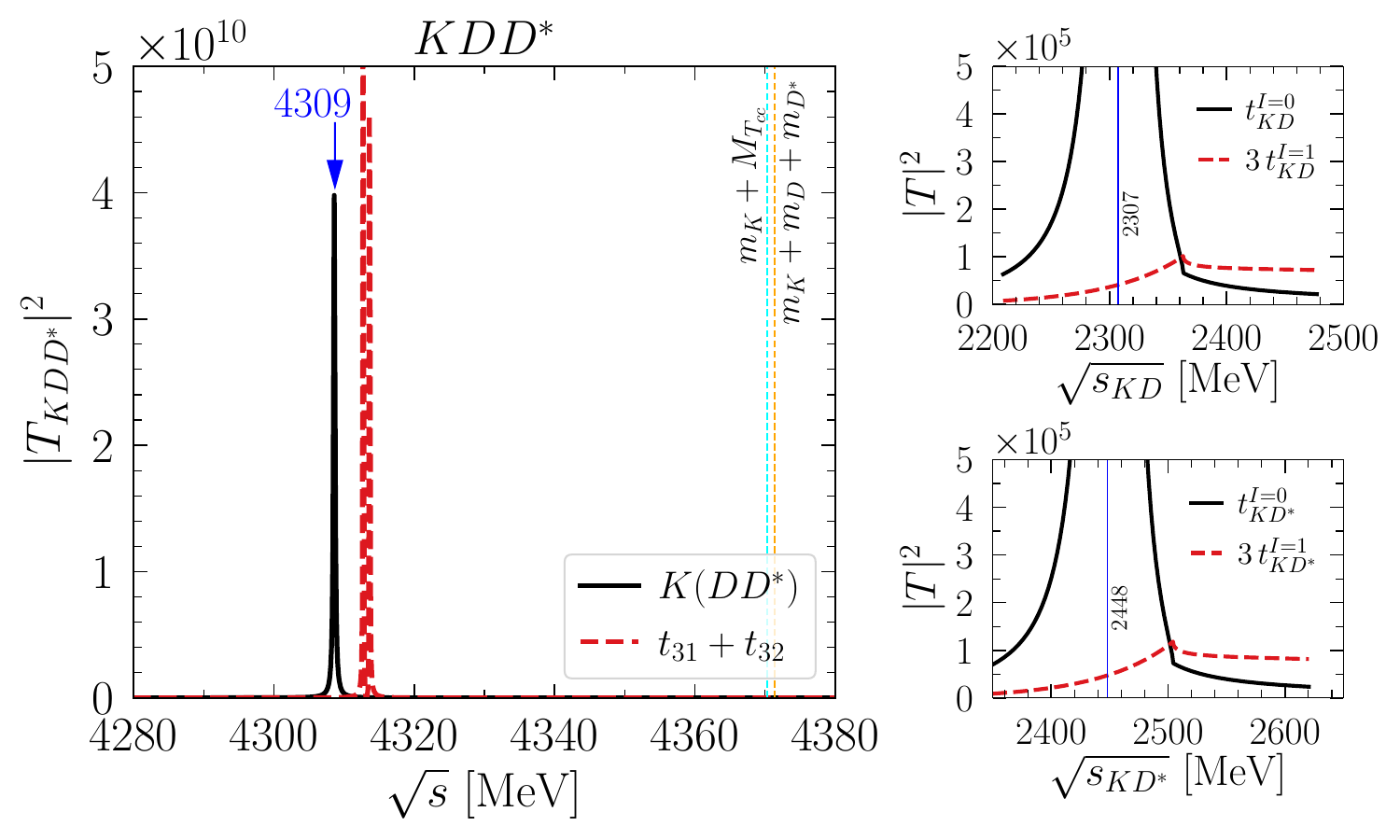}
  \caption{Left panel: Modulus squared of the total scattering amplitude for the $KDD^*$ system with isospin $I=1/2$. The thresholds for the $KT_{cc}$ and $KDD^*$ systems are indicated by vertical lines. Right panel: Modulus squared of the two-body scattering amplitudes for $KD$ and $KD^*$ with isospin $I=0,\, 1$. The (blue) vertical lines denote the values of $\sqrt{s_{KD}}$ and $\sqrt{s_{KD^*}}$ for $\sqrt{s}=4309$, which is the mass of the $KDD^*$ bound state found.}
  \label{Fig:T2_KTcc}
\end{figure*}

\begin{figure*}[t]
  \includegraphics[width=0.7\textwidth]{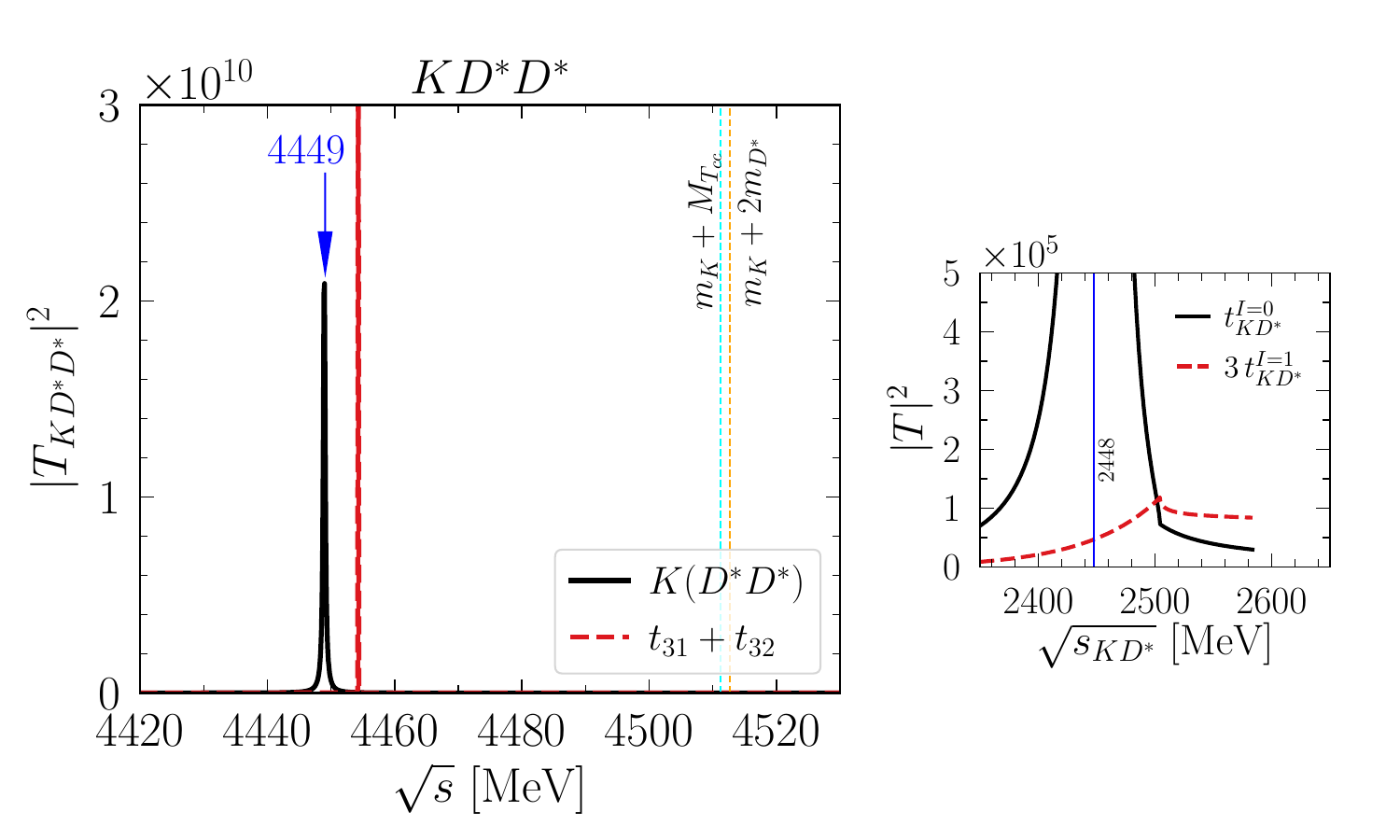}
  \caption{Left panel: Modulus squared of the total scattering amplitude for the $KD^*D^*$  system with isospin $I=1/2$. The thresholds related to the $KT_{cc}$ and $KD^*D^*$ systems are indicated by vertical lines. Right panel: Modulus squared of the two-body scattering amplitudes for the $KD^*$ system with isospin $I=0,\, 1$. The (blue) vertical line marks the position $\sqrt{s_{KD^*}}=2448$ MeV, which is the value of that invariant mass for a center-of-mass energy of $\sqrt{s}=4449$ MeV, i.e., the mass associated with the $KD^*D^*$ bound state formed.}
  \label{Fig:T2_KDxDx}
\end{figure*}

\subsection{Green function for $K$ propagating in the cluster $R_{12}$}
The loop function $G_0$ in Eq.~\eqref{Eq:FCA}, which describes the propagation of the $K$ meson in the $R_{12}$ cluster, can be expressed as~\cite{Roca:2010tf,Yamagata-Sekihara:2010kpd} 
\begin{equation}\label{Eq:G0exp}
	G_0(s) = \frac{1}{2M_{R}} \int \frac{d^3 \bm{q}}{(2\pi)^3} \frac{F_{R}(\bm{q}^2)}{(q^0)^2 - \bm{q}^2 - m_{K}^2 + i\epsilon},
\end{equation}
where $M_R$ is the mass of the $R_{12}$ cluster, and  $q^0$ is the on-shell energy of the $K$ meson in the cluster rest frame: 
\begin{equation}
q^0=\frac{s-m_K^2-M_{R}^2}{2M_R}\, ,	
\end{equation}
In Eq.~(\ref{Eq:G0exp}), $F_{R}(q)$ is the form factor of $R_{12}$ as a cluster of two hadrons in s-wave, which is related to the corresponding wavefunction in the coordinate representation via a Fourier transformation~\cite{Roca:2010tf,Yamagata-Sekihara:2010muv,Yamagata-Sekihara:2010kpd,MartinezTorres:2020hus}: 
\begin{equation}\label{Eq:FR}
\begin{aligned}
	  F_{R}(q) &= \frac{1}{\mathcal{N}} \int_{\{|\bm{p}|, |\bm{p}-\bm{q}|<\Lambda_{R}\}} d^3 \bm{p} \, f_R(\bm{p})\, f_{R}(\bm{p}-\bm{q}), \\
	  \mathcal{N} & =\int_{|\bm{p}|<\Lambda_{R}} d^3 \bm{p} \, \bigl[f_R(\bm{p})\bigr]^2\, ,\\
	  f_{R}(\bm{p}) &= 
  \frac{1}{2\,\omega_{1}(\bm{p})}\frac{1}{2\,\omega_{2}(\bm{p})} \frac{1}{M_{R}-\omega_{1}(\bm{p})-\omega_{2}(\bm{p})} \, ,
\end{aligned}
\end{equation}
with the normalization being $F_R(q=0)=1$. In Eq.~\eqref{Eq:FR}, the particle energies $\omega_{1(2)}$ are given by $\omega_{1}(\bm{p})=\sqrt{m_{1}^2+\bm{p}^2}$, $\omega_{2}(\bm{p})=\sqrt{m_{2}^2+\bm{p}^2}$. 
 Since the form factor $F_{R}(q)$ is related to the wave function of the $R_{12}$ cluster, the upper integration limit $\Lambda_{R}$ in Eq.~\eqref{Eq:FR} corresponds to the value of the momentum cutoff considered in the study where the generation of the cluster from the interaction of particles $P_1$ and $P_2$ is found when solving the Bethe-Salpeter equation, where two-hadron loop functions need to be regularized~\cite{MartinezTorres:2020hus}. For the present systems under study, i.e., $KDD^*$ and $KD^*D^*$, we list in Table~\ref{Tab:MRLambR} the values considered for the cluster masses $M_R$ of $DD^*$ and $D^*D^*$ and the corresponding cutoff $\Lambda_R$, values which are based on the results found in Refs.~\cite{Feijoo:2021ppq,Dai:2021vgf}. The resulting form factors $F_{R}(q)$ are shown in Fig.~\ref{Fig:FormFactor} for the $DD^*$ and $D^*D^*$ clusters. As can be seen, they start from $1$ and gradually decrease to $0$ as $q$ increases. Also, for $q\sim 2 \Lambda_{R}$, both form factors are exactly zero due to the integral bound $|\bm{p}-\bm{q}|< \Lambda_R$ present in Eq.~\eqref{Eq:FR}.

Furthermore, in Fig.~\ref{Fig:G0} we present the real and imaginary parts of the Green function $G_0(s)$ for the case of the $KDD^*$ and $KD^*D^*$ systems. As can be seen, it presents the typical feature of a two-particle loop function, i.e., a cusp at the $m_3+M_R$ threshold for its real part and a non-zero imaginary part for energies above that threshold, which is $\sqrt{s}=4370.5$ MeV for the $KDD^*$ system, and $\sqrt{s}=4511.2$ MeV for the $KD^*D^*$ case.

\section{Results and discussion}\label{Sec:Results}
In this section, we present the three-body scattering amplitudes for the $KDD^*$ and $KD^*D^*$ systems for the total isospin $I=1/2$ and spin-parity $J^P=1^-$. Using the expressions for $t_{31}$ and $t_{32}$ in Eqs.~\eqref{Eq:t3132inKDDx}, \eqref{Eq:t3132inKDxDx}, along with the weight factors in Eq.~\eqref{Eq:WeightFactor}, we can determine $T_\mathrm{tol}$ from Eq.~\eqref{Eq:Ttol} for the $K(DD^*)$ and $K(D^*D^*)$ systems, as a function of $\sqrt{s}$. 

In Fig.~\ref{Fig:T2_KTcc}, we present the modulus squared $|T_\text{tot}|^2$ for the transition $KT_{cc} \to K T_{cc}$ as a function of $\sqrt{s}$. A narrow peak around $4309$ MeV is observed, indicating the existence of a $KDD^*$ bound state with a binding energy $B_{KDD^*}\simeq 62$ MeV. This state is dynamically generated through the non-perturbative iteration of the single scattering amplitudes $t_{31}$ and $t_{32}$, as given in Eq.~\eqref{Eq:Ttol}. To show this explicitly, we also present in Fig.~\ref{Fig:T2_KTcc} the result obtained from $|t_{31}+t_{32}|^2$, where $t_{31}+t_{32}$ represents the input for Eq.~\eqref{Eq:Ttol}. As can be seen, two nearby peaks are observed at $\sqrt{s}=4312.7$ MeV and $4313.6$ MeV, which are a reflection of the two-body resonances $D_{s0}^*(2317)$ and $D_{s1}(2460)$ generated in the two-body systems. These two peaks, by summing all the rescattering contributions of particle 3 with particles 1 and 2, ultimately lead to the formation of the $KDD^*$ bound state at $\sqrt{s}=4309$ MeV. 

For a better understanding of the dynamics involved in the generation of the aforementioned three-body state, we show the modulus squared of the two-body scattering amplitudes $t_{KD}$ and $t_{KD^*}$ with $I=0,\,1$ as a function of the corresponding two-body invariant masses on the right panel of Fig.~\ref{Fig:T2_KTcc}. At the energy of the three-body bound state, i.e., $\sqrt{s}=4309$ MeV, the corresponding energies of the $KD$ and $KD^*$ subsystems are $\sqrt{s_{KD}}=2307$ MeV and $\sqrt{s_{KD^*}}=2448$ MeV [see Eq.~\eqref{Eq:s31s32}]. As can be seen in Fig.~\ref{Fig:T2_KTcc}, this is precisely the energy region in which $D^*_{s0}(2317)$ and $D_{s1}(2460)$ are generated from, the $KD$ and $KD^*$ interactions, respectively, in isospin $I=0$ and the latter amplitudes are much bigger than their $I=1$ counterparts. In this way, we can conclude that the $D^*_{s0}(2317)$ and the $D_{s1}(2460)$ are generated in the (31) and (32) subsystems, while $T_{cc}$ is formed in the $D D^*$ subsystems. All these attractive interactions produce a state which is $\sim 4$ MeV away from the position of the peaks seen in $|t_{31}+t_{32}|^2$.

Next, we focus on the $K(D^*D^*)$ system, for which we show the modulus squared of the $T$-matrix in the isospin $I=1/2$ and spin-parity $J^P=1^-$ as a function of the center-of-mass energy in Fig.~\ref{Fig:T2_KDxDx}. As can be seen, there exists a bound state with a mass of $4449$ MeV, which is $\simeq 64$ MeV below the three-body threshold $m_{K}+2m_{D^*}$. 
By considering $|t_{31}+t_{32}|^2$ alone, we simply find the reflection of the $D_{s1}(2460)$ state originated from the $KD^*$ interaction, whose mass has a related $\sqrt{s}$ value of $4454$~MeV. It should be recalled that the two-body states are generated in the presence of a third particle, and are expected to appear at a total energy which depends on the mass of the third particle and the peak position in the invariant mass of the resonating pair. Only when all contributions represented in Fig.~\ref{Fig:FCA} are implemented via the resolution of Eq.~(\ref{Eq:FCA}), a three-body state, which is $\sim 5$ MeV away from the position of the peaks in $|t_{31}+t_{32}|^2$, is observed. Furthermore, we present $|t_{KD^*}|^2$ for $I=0,\, 1$ in the right panel of Fig.~\ref{Fig:T2_KDxDx} as a function of the invariant mass of the system. Similarly to the $KDD^*$ system, the $KD^*$ interaction with $I=0$ is much stronger than that with $I=1$ as a consequence of the generation of $D_{s1}(2460)$. As can be also seen, for the $\sqrt{s}$ value at which the three-body state is found, the $KD^*$ interaction generates $D_{s1}(2460)$, which together with the formation of a $T_{cc}$ like state with spin 1 in the $D^*D^*$ system~\cite{Dai:2021vgf}, it produces such a strong attraction which binds the three-body system.

We summarize our results for the masses and binding energies of the states obtained in the $KDD^*$ and $KD^*D^*$ systems in Table~\ref{Tab:MassCompare}, and compare them with the previous calculations available for the $KDD^*$ system with the same quantum numbers~\cite{Ma:2017ery,Zhang:2024yfj}. Our result for $K^*_{cc}(4309)$ is located in between the two previous estimations. Such uncertainties in the masses are common among hadronic models based on different methods to study the interactions between the particles constituting the system. To be more specific, in Ref.~\cite{Ma:2017ery}, the one-pion-exchange potentials among the two-body subsystems present in the $KDD^*$ system are used within the Born-Oppenheimer approximation to determine the wave function of the system and find the corresponding binding energies. 
In Ref.~\cite{Zhang:2024yfj}, the authors consider chiral interactions for $KD$ and $KD^*$ beyond the leading order, along with the lowest order $DD^*$ potential. They obtain a $KDD^*$ bound state with a binding energy ranging from $44$ to $84$ MeV by solving the three-body system within the  lattice effective field theory framework~\cite{Zhang:2024yfj}. 

In spite of the uncertainties inherent to the consideration of different inputs and methods employed to determine the three-body $T$-matrix,
there is no doubt that the generation of a three-body state with a mass $\simeq 4300$ MeV is obtained with exotic quantum numbers of strangeness $+1$ and charm $+2$, and $J^P=1^-$. We also predict the existence of a $J^P=1^-$ state with strangeness $+1$, charm $+2$, and a higher mass as a consequence of the $KD^*D^*$ dynamics. 
 It should be noted here that the $KD^*D^*$ dynamics produces a state slightly more bound than the one in $KDD^*$, a result which is mainly due to the fact that the input $KD^*$ interaction is slightly more attractive than the $KD$ interaction. This can be deduced from the binding energy of the bound states obtained ($\sim 52$ MeV in the case of the $KD^*$ system and $\sim 50$ MeV for the $KD$ system). Furthermore, the $D^*D^*$ interaction considered is also more attractive than that of $DD^*$, as can be deduced from the results shown in Table~\ref{Tab:MRLambR}. We should, however, mention that  a difference of the order of 2 MeV in the binding energies of the states is compatible with the expected uncertainties in the inputs, making it difficult to affirm with certainty that $KD^*D^*$ interaction is more attractive than that of $KDD^*$. 

Future experimental searches will be relevant for the confirmation of these states. Considering the internal structures obtained for the aforementioned states, $K_{cc}^*(4309)$ can decay, for example, into two particle final states like $DD_s$ and $D^*D_s^*$ via a mechanism involving triangular loops. Similarly, $K_{cc}^*(4449)$ can decay into two-body channels like $D^*D_s$ and $DD_s^*$. The corresponding calculation of these partial decay widths is currently in progress. Feasible experimental searches of these $K_{cc}^*$ states could involve the reconstruction of the $D^{(*)}D_s^{(*)}$ invariant mass distribution from their production in $\Upsilon(1S,2S)$ inclusive decays or from direct production in $e^+e^-$ collisions, similar to the study in Refs.~\cite{Belle:2020xca,Belle:2021kub}.  The recent progress in the studies of doubly heavy-flavored hadrons within lattice QCD~\cite{Meng:2024kkp,Junnarkar:2024kwd,Francis:2024fwf,Whyte:2024ihh,Radhakrishnan:2024ihu,Collins:2024sfi,Padmanath:2023rdu,Aoki:2023nzp}  shows that the exploration of the existence of $K_{cc}^*$ is also plausible in such a framework.

\begin{table}[t]
\centering
\caption{Masses and binding energies of the $KDD^*$ and $KD^*D^*$ bound states with $J^P=1^-$, and a comparison with previous calculations.}
\label{Tab:MassCompare}
\begin{tabular}{c| c c | c c}
\hline\hline 
 & \multicolumn{2}{ c| }{$KDD^*$} & \multicolumn{2}{c}{$KD^*D^*$} \\ \hline
  & $M$~[MeV] & $B$~[MeV] & $M$~[MeV] & $B$~[MeV]  \\ \hline
This work & $4309$ & $62$   & $4449$ & $64$ \\ 
\hline
Ref.~\cite{Zhang:2024yfj} & $4292(4)$ & $79(4)$ & $-$ & $-$ \\ \hline
Ref.~\cite{Ma:2017ery}    & $4318(4)$ & $54(4)$  & $-$ & $-$ \\ 
\hline\hline 
\end{tabular}
\end{table}

\section{Summary} \label{Sec:Summary}
In this work, we have investigated the three-body systems $KDD^*$ and $KD^*D^*$ to search for potential exotic states with open strange and double charm flavor. 
Inspired by the recent observation of the $T_{cc}$ state by the LHCb Collaboration, we consider the scattering of a Kaon with the $D D^*$ and $D^*D^*$ systems, where $DD^*$ clusters as $T_{cc}$, while $D^*D^*$ forms a $T_{cc}$ like state, as shown in Refs.~\cite{Feijoo:2021ppq,Dai:2021vgf}, with both clusters having $I=0$ and $J^P=1^+$. Besides this, the two-body $KD$ and $KD^*$ amplitudes in the $I=0,\,1$ sectors are well constrained from the heavy-quark and chiral symmetries, giving rise to the generation of the $D_{s0}^*(2317)$ and $D_{s1}(2460)$ resonances. Through repeatedly scattering of the light $K$ meson off the particles forming the heavy clusters, we evaluated the three-body amplitudes by using the fixed-center approximation to the Faddeev equations. We find two bound states, named $K^*_{cc}(4309)$ and $K^*_{cc}(4449)$, with quantum numbers $I(J^P)=1/2(1^-)$ for the $KDD^*$ and $KD^*D^*$ systems, respectively. The corresponding binding energies are of the order of $60$ MeV. We hope this work encourages future experimental searches for these hexaquark mesons containing $ccs$ open flavors. 

\acknowledgements
This work was partly supported by the Deutsche Forschungsgemeinschaft (DFG, German Research Foundation), through the Research Unit (Projektnummer 458854507—FOR 5327), and the Cluster of Excellence (PRISMA$^+$ EXC 2118/1) within the German Excellence Strategy (Projektnummer 39083149), CNPq ( K.P.K: Grants No. 407437/ 2023-1 and No. 306461/2023-4; A.M.T: Grant No. 304510/2023-8), and FAPESP (K.P.K.: Grant Number 2022/08347-9; A. M. T.: Grant number 2023/01182-7).

\bibliographystyle{apsrev4-2}
\bibliography{ref_KTcc.bib}

\begin{thebibliography}{72}%
\makeatletter
\providecommand \@ifxundefined [1]{%
 \@ifx{#1\undefined}
}%
\providecommand \@ifnum [1]{%
 \ifnum #1\expandafter \@firstoftwo
 \else \expandafter \@secondoftwo
 \fi
}%
\providecommand \@ifx [1]{%
 \ifx #1\expandafter \@firstoftwo
 \else \expandafter \@secondoftwo
 \fi
}%
\providecommand \natexlab [1]{#1}%
\providecommand \enquote  [1]{``#1''}%
\providecommand \bibnamefont  [1]{#1}%
\providecommand \bibfnamefont [1]{#1}%
\providecommand \citenamefont [1]{#1}%
\providecommand \href@noop [0]{\@secondoftwo}%
\providecommand \href [0]{\begingroup \@sanitize@url \@href}%
\providecommand \@href[1]{\@@startlink{#1}\@@href}%
\providecommand \@@href[1]{\endgroup#1\@@endlink}%
\providecommand \@sanitize@url [0]{\catcode `\\12\catcode `\$12\catcode `\&12\catcode `\#12\catcode `\^12\catcode `\_12\catcode `\%12\relax}%
\providecommand \@@startlink[1]{}%
\providecommand \@@endlink[0]{}%
\providecommand \url  [0]{\begingroup\@sanitize@url \@url }%
\providecommand \@url [1]{\endgroup\@href {#1}{\urlprefix }}%
\providecommand \urlprefix  [0]{URL }%
\providecommand \Eprint [0]{\href }%
\providecommand \doibase [0]{https://doi.org/}%
\providecommand \selectlanguage [0]{\@gobble}%
\providecommand \bibinfo  [0]{\@secondoftwo}%
\providecommand \bibfield  [0]{\@secondoftwo}%
\providecommand \translation [1]{[#1]}%
\providecommand \BibitemOpen [0]{}%
\providecommand \bibitemStop [0]{}%
\providecommand \bibitemNoStop [0]{.\EOS\space}%
\providecommand \EOS [0]{\spacefactor3000\relax}%
\providecommand \BibitemShut  [1]{\csname bibitem#1\endcsname}%
\let\auto@bib@innerbib\@empty
\bibitem [{\citenamefont {Choi}\ \emph {et~al.}(2003)\citenamefont {Choi} \emph {et~al.}}]{Belle:2003nnu}%
  \BibitemOpen
  \bibfield  {author} {\bibinfo {author} {\bibfnamefont {S.~K.}\ \bibnamefont {Choi}} \emph {et~al.} (\bibinfo {collaboration} {Belle}),\ }\href {https://doi.org/10.1103/PhysRevLett.91.262001} {\bibfield  {journal} {\bibinfo  {journal} {Phys. Rev. Lett.}\ }\textbf {\bibinfo {volume} {91}},\ \bibinfo {pages} {262001} (\bibinfo {year} {2003})},\ \Eprint {https://arxiv.org/abs/hep-ex/0309032} {arXiv:hep-ex/0309032} \BibitemShut {NoStop}%
\bibitem [{\citenamefont {Ablikim}\ \emph {et~al.}(2013)\citenamefont {Ablikim} \emph {et~al.}}]{BESIII:2013ris}%
  \BibitemOpen
  \bibfield  {author} {\bibinfo {author} {\bibfnamefont {M.}~\bibnamefont {Ablikim}} \emph {et~al.} (\bibinfo {collaboration} {BESIII}),\ }\href {https://doi.org/10.1103/PhysRevLett.110.252001} {\bibfield  {journal} {\bibinfo  {journal} {Phys. Rev. Lett.}\ }\textbf {\bibinfo {volume} {110}},\ \bibinfo {pages} {252001} (\bibinfo {year} {2013})},\ \Eprint {https://arxiv.org/abs/1303.5949} {arXiv:1303.5949 [hep-ex]} \BibitemShut {NoStop}%
\bibitem [{\citenamefont {Liu}\ \emph {et~al.}(2013)\citenamefont {Liu} \emph {et~al.}}]{Belle:2013yex}%
  \BibitemOpen
  \bibfield  {author} {\bibinfo {author} {\bibfnamefont {Z.~Q.}\ \bibnamefont {Liu}} \emph {et~al.} (\bibinfo {collaboration} {Belle}),\ }\href {https://doi.org/10.1103/PhysRevLett.110.252002} {\bibfield  {journal} {\bibinfo  {journal} {Phys. Rev. Lett.}\ }\textbf {\bibinfo {volume} {110}},\ \bibinfo {pages} {252002} (\bibinfo {year} {2013})},\ \Eprint {https://arxiv.org/abs/1304.0121} {arXiv:1304.0121 [hep-ex]} \BibitemShut {NoStop}%
\bibitem [{\citenamefont {Aaij}\ \emph {et~al.}(2022)\citenamefont {Aaij} \emph {et~al.}}]{LHCb:2021auc}%
  \BibitemOpen
  \bibfield  {author} {\bibinfo {author} {\bibfnamefont {R.}~\bibnamefont {Aaij}} \emph {et~al.} (\bibinfo {collaboration} {LHCb}),\ }\href {https://doi.org/10.1038/s41467-022-30206-w} {\bibfield  {journal} {\bibinfo  {journal} {Nature Commun.}\ }\textbf {\bibinfo {volume} {13}},\ \bibinfo {pages} {3351} (\bibinfo {year} {2022})},\ \Eprint {https://arxiv.org/abs/2109.01056} {arXiv:2109.01056 [hep-ex]} \BibitemShut {NoStop}%
\bibitem [{\citenamefont {Aubert}\ \emph {et~al.}(2003)\citenamefont {Aubert} \emph {et~al.}}]{BaBar:2003oey}%
  \BibitemOpen
  \bibfield  {author} {\bibinfo {author} {\bibfnamefont {B.}~\bibnamefont {Aubert}} \emph {et~al.} (\bibinfo {collaboration} {BaBar}),\ }\href {https://doi.org/10.1103/PhysRevLett.90.242001} {\bibfield  {journal} {\bibinfo  {journal} {Phys. Rev. Lett.}\ }\textbf {\bibinfo {volume} {90}},\ \bibinfo {pages} {242001} (\bibinfo {year} {2003})},\ \Eprint {https://arxiv.org/abs/hep-ex/0304021} {arXiv:hep-ex/0304021} \BibitemShut {NoStop}%
\bibitem [{\citenamefont {Besson}\ \emph {et~al.}(2003)\citenamefont {Besson} \emph {et~al.}}]{CLEO:2003ggt}%
  \BibitemOpen
  \bibfield  {author} {\bibinfo {author} {\bibfnamefont {D.}~\bibnamefont {Besson}} \emph {et~al.} (\bibinfo {collaboration} {CLEO}),\ }\href {https://doi.org/10.1103/PhysRevD.68.032002} {\bibfield  {journal} {\bibinfo  {journal} {Phys. Rev. D}\ }\textbf {\bibinfo {volume} {68}},\ \bibinfo {pages} {032002} (\bibinfo {year} {2003})},\ \Eprint {https://arxiv.org/abs/hep-ex/0305100} {arXiv:hep-ex/0305100} \BibitemShut {NoStop}%
\bibitem [{\citenamefont {Aaij}\ \emph {et~al.}(2020)\citenamefont {Aaij} \emph {et~al.}}]{LHCb:2020bwg}%
  \BibitemOpen
  \bibfield  {author} {\bibinfo {author} {\bibfnamefont {R.}~\bibnamefont {Aaij}} \emph {et~al.} (\bibinfo {collaboration} {LHCb}),\ }\href {https://doi.org/10.1016/j.scib.2020.08.032} {\bibfield  {journal} {\bibinfo  {journal} {Sci. Bull.}\ }\textbf {\bibinfo {volume} {65}},\ \bibinfo {pages} {1983} (\bibinfo {year} {2020})},\ \Eprint {https://arxiv.org/abs/2006.16957} {arXiv:2006.16957 [hep-ex]} \BibitemShut {NoStop}%
\bibitem [{\citenamefont {Aaij}\ \emph {et~al.}(2015)\citenamefont {Aaij} \emph {et~al.}}]{LHCb:2015yax}%
  \BibitemOpen
  \bibfield  {author} {\bibinfo {author} {\bibfnamefont {R.}~\bibnamefont {Aaij}} \emph {et~al.} (\bibinfo {collaboration} {LHCb}),\ }\href {https://doi.org/10.1103/PhysRevLett.115.072001} {\bibfield  {journal} {\bibinfo  {journal} {Phys. Rev. Lett.}\ }\textbf {\bibinfo {volume} {115}},\ \bibinfo {pages} {072001} (\bibinfo {year} {2015})},\ \Eprint {https://arxiv.org/abs/1507.03414} {arXiv:1507.03414 [hep-ex]} \BibitemShut {NoStop}%
\bibitem [{\citenamefont {Aaij}\ \emph {et~al.}(2019)\citenamefont {Aaij} \emph {et~al.}}]{LHCb:2019kea}%
  \BibitemOpen
  \bibfield  {author} {\bibinfo {author} {\bibfnamefont {R.}~\bibnamefont {Aaij}} \emph {et~al.} (\bibinfo {collaboration} {LHCb}),\ }\href {https://doi.org/10.1103/PhysRevLett.122.222001} {\bibfield  {journal} {\bibinfo  {journal} {Phys. Rev. Lett.}\ }\textbf {\bibinfo {volume} {122}},\ \bibinfo {pages} {222001} (\bibinfo {year} {2019})},\ \Eprint {https://arxiv.org/abs/1904.03947} {arXiv:1904.03947 [hep-ex]} \BibitemShut {NoStop}%
\bibitem [{\citenamefont {Olsen}\ \emph {et~al.}(2018)\citenamefont {Olsen}, \citenamefont {Skwarnicki},\ and\ \citenamefont {Zieminska}}]{Olsen:2017bmm}%
  \BibitemOpen
  \bibfield  {author} {\bibinfo {author} {\bibfnamefont {S.~L.}\ \bibnamefont {Olsen}}, \bibinfo {author} {\bibfnamefont {T.}~\bibnamefont {Skwarnicki}},\ and\ \bibinfo {author} {\bibfnamefont {D.}~\bibnamefont {Zieminska}},\ }\href {https://doi.org/10.1103/RevModPhys.90.015003} {\bibfield  {journal} {\bibinfo  {journal} {Rev. Mod. Phys.}\ }\textbf {\bibinfo {volume} {90}},\ \bibinfo {pages} {015003} (\bibinfo {year} {2018})},\ \Eprint {https://arxiv.org/abs/1708.04012} {arXiv:1708.04012 [hep-ph]} \BibitemShut {NoStop}%
\bibitem [{\citenamefont {Brambilla}\ \emph {et~al.}(2020)\citenamefont {Brambilla}, \citenamefont {Eidelman}, \citenamefont {Hanhart}, \citenamefont {Nefediev}, \citenamefont {Shen}, \citenamefont {Thomas}, \citenamefont {Vairo},\ and\ \citenamefont {Yuan}}]{Brambilla:2019esw}%
  \BibitemOpen
  \bibfield  {author} {\bibinfo {author} {\bibfnamefont {N.}~\bibnamefont {Brambilla}}, \bibinfo {author} {\bibfnamefont {S.}~\bibnamefont {Eidelman}}, \bibinfo {author} {\bibfnamefont {C.}~\bibnamefont {Hanhart}}, \bibinfo {author} {\bibfnamefont {A.}~\bibnamefont {Nefediev}}, \bibinfo {author} {\bibfnamefont {C.-P.}\ \bibnamefont {Shen}}, \bibinfo {author} {\bibfnamefont {C.~E.}\ \bibnamefont {Thomas}}, \bibinfo {author} {\bibfnamefont {A.}~\bibnamefont {Vairo}},\ and\ \bibinfo {author} {\bibfnamefont {C.-Z.}\ \bibnamefont {Yuan}},\ }\href {https://doi.org/10.1016/j.physrep.2020.05.001} {\bibfield  {journal} {\bibinfo  {journal} {Phys. Rept.}\ }\textbf {\bibinfo {volume} {873}},\ \bibinfo {pages} {1} (\bibinfo {year} {2020})},\ \Eprint {https://arxiv.org/abs/1907.07583} {arXiv:1907.07583 [hep-ex]} \BibitemShut {NoStop}%
\bibitem [{\citenamefont {Meng}\ \emph {et~al.}(2023)\citenamefont {Meng}, \citenamefont {Wang}, \citenamefont {Wang},\ and\ \citenamefont {Zhu}}]{Meng:2022ozq}%
  \BibitemOpen
  \bibfield  {author} {\bibinfo {author} {\bibfnamefont {L.}~\bibnamefont {Meng}}, \bibinfo {author} {\bibfnamefont {B.}~\bibnamefont {Wang}}, \bibinfo {author} {\bibfnamefont {G.-J.}\ \bibnamefont {Wang}},\ and\ \bibinfo {author} {\bibfnamefont {S.-L.}\ \bibnamefont {Zhu}},\ }\href {https://doi.org/10.1016/j.physrep.2023.04.003} {\bibfield  {journal} {\bibinfo  {journal} {Phys. Rept.}\ }\textbf {\bibinfo {volume} {1019}},\ \bibinfo {pages} {2266} (\bibinfo {year} {2023})},\ \Eprint {https://arxiv.org/abs/2204.08716} {arXiv:2204.08716 [hep-ph]} \BibitemShut {NoStop}%
\bibitem [{\citenamefont {Mai}\ \emph {et~al.}(2023)\citenamefont {Mai}, \citenamefont {Mei{\ss}ner},\ and\ \citenamefont {Urbach}}]{Mai:2022eur}%
  \BibitemOpen
  \bibfield  {author} {\bibinfo {author} {\bibfnamefont {M.}~\bibnamefont {Mai}}, \bibinfo {author} {\bibfnamefont {U.-G.}\ \bibnamefont {Mei{\ss}ner}},\ and\ \bibinfo {author} {\bibfnamefont {C.}~\bibnamefont {Urbach}},\ }\href {https://doi.org/10.1016/j.physrep.2022.11.005} {\bibfield  {journal} {\bibinfo  {journal} {Phys. Rept.}\ }\bibinfo {series} {Towards a Theory of Hadron Resonances},\ \textbf {\bibinfo {volume} {1001}},\ \bibinfo {pages} {2248} (\bibinfo {year} {2023})},\ \Eprint {https://arxiv.org/abs/2206.01477} {arXiv:2206.01477 [hep-ph]} \BibitemShut {NoStop}%
\bibitem [{\citenamefont {Chen}\ \emph {et~al.}(2022)\citenamefont {Chen}, \citenamefont {Chen}, \citenamefont {Liu}, \citenamefont {Liu},\ and\ \citenamefont {Zhu}}]{Chen:2022asf}%
  \BibitemOpen
  \bibfield  {author} {\bibinfo {author} {\bibfnamefont {H.-X.}\ \bibnamefont {Chen}}, \bibinfo {author} {\bibfnamefont {W.}~\bibnamefont {Chen}}, \bibinfo {author} {\bibfnamefont {X.}~\bibnamefont {Liu}}, \bibinfo {author} {\bibfnamefont {Y.-R.}\ \bibnamefont {Liu}},\ and\ \bibinfo {author} {\bibfnamefont {S.-L.}\ \bibnamefont {Zhu}},\ }\href {https://doi.org/10.1088/1361-6633/aca3b6} {\bibfield  {journal} {\bibinfo  {journal} {Rept. Prog. Phys.}\ }\textbf {\bibinfo {volume} {86}},\ \bibinfo {pages} {026201} (\bibinfo {year} {2022})},\ \Eprint {https://arxiv.org/abs/2204.02649} {arXiv:2204.02649 [hep-ph]} \BibitemShut {NoStop}%
\bibitem [{\citenamefont {Liu}\ \emph {et~al.}(2019)\citenamefont {Liu}, \citenamefont {Chen}, \citenamefont {Chen}, \citenamefont {Liu},\ and\ \citenamefont {Zhu}}]{Liu:2019zoy}%
  \BibitemOpen
  \bibfield  {author} {\bibinfo {author} {\bibfnamefont {Y.-R.}\ \bibnamefont {Liu}}, \bibinfo {author} {\bibfnamefont {H.-X.}\ \bibnamefont {Chen}}, \bibinfo {author} {\bibfnamefont {W.}~\bibnamefont {Chen}}, \bibinfo {author} {\bibfnamefont {X.}~\bibnamefont {Liu}},\ and\ \bibinfo {author} {\bibfnamefont {S.-L.}\ \bibnamefont {Zhu}},\ }\href {https://doi.org/10.1016/j.ppnp.2019.04.003} {\bibfield  {journal} {\bibinfo  {journal} {Prog. Part. Nucl. Phys.}\ }\textbf {\bibinfo {volume} {107}},\ \bibinfo {pages} {237} (\bibinfo {year} {2019})},\ \Eprint {https://arxiv.org/abs/1903.11976} {arXiv:1903.11976 [hep-ph]} \BibitemShut {NoStop}%
\bibitem [{\citenamefont {Guo}\ \emph {et~al.}(2018)\citenamefont {Guo}, \citenamefont {Hanhart}, \citenamefont {Mei{\ss}ner}, \citenamefont {Wang}, \citenamefont {Zhao},\ and\ \citenamefont {Zou}}]{Guo:2017jvc}%
  \BibitemOpen
  \bibfield  {author} {\bibinfo {author} {\bibfnamefont {F.-K.}\ \bibnamefont {Guo}}, \bibinfo {author} {\bibfnamefont {C.}~\bibnamefont {Hanhart}}, \bibinfo {author} {\bibfnamefont {U.-G.}\ \bibnamefont {Mei{\ss}ner}}, \bibinfo {author} {\bibfnamefont {Q.}~\bibnamefont {Wang}}, \bibinfo {author} {\bibfnamefont {Q.}~\bibnamefont {Zhao}},\ and\ \bibinfo {author} {\bibfnamefont {B.-S.}\ \bibnamefont {Zou}},\ }\href {https://doi.org/10.1103/RevModPhys.90.015004} {\bibfield  {journal} {\bibinfo  {journal} {Rev. Mod. Phys.}\ }\textbf {\bibinfo {volume} {90}},\ \bibinfo {pages} {015004} (\bibinfo {year} {2018})},\ \Eprint {https://arxiv.org/abs/1705.00141} {arXiv:1705.00141 [hep-ph]} \BibitemShut {NoStop}%
\bibitem [{\citenamefont {Gamermann}\ and\ \citenamefont {Oset}(2007)}]{Gamermann:2007fi}%
  \BibitemOpen
  \bibfield  {author} {\bibinfo {author} {\bibfnamefont {D.}~\bibnamefont {Gamermann}}\ and\ \bibinfo {author} {\bibfnamefont {E.}~\bibnamefont {Oset}},\ }\href {https://doi.org/10.1140/epja/i2007-10435-1} {\bibfield  {journal} {\bibinfo  {journal} {Eur. Phys. J. A}\ }\textbf {\bibinfo {volume} {33}},\ \bibinfo {pages} {119} (\bibinfo {year} {2007})},\ \Eprint {https://arxiv.org/abs/0704.2314} {arXiv:0704.2314 [hep-ph]} \BibitemShut {NoStop}%
\bibitem [{\citenamefont {Guo}\ \emph {et~al.}(2013)\citenamefont {Guo}, \citenamefont {Hanhart}, \citenamefont {Mei\ss{}ner}, \citenamefont {Wang},\ and\ \citenamefont {Zhao}}]{Guo:2013zbw}%
  \BibitemOpen
  \bibfield  {author} {\bibinfo {author} {\bibfnamefont {F.-K.}\ \bibnamefont {Guo}}, \bibinfo {author} {\bibfnamefont {C.}~\bibnamefont {Hanhart}}, \bibinfo {author} {\bibfnamefont {U.-G.}\ \bibnamefont {Mei\ss{}ner}}, \bibinfo {author} {\bibfnamefont {Q.}~\bibnamefont {Wang}},\ and\ \bibinfo {author} {\bibfnamefont {Q.}~\bibnamefont {Zhao}},\ }\href {https://doi.org/10.1016/j.physletb.2013.06.053} {\bibfield  {journal} {\bibinfo  {journal} {Phys. Lett. B}\ }\textbf {\bibinfo {volume} {725}},\ \bibinfo {pages} {127} (\bibinfo {year} {2013})},\ \Eprint {https://arxiv.org/abs/1306.3096} {arXiv:1306.3096 [hep-ph]} \BibitemShut {NoStop}%
\bibitem [{\citenamefont {Aceti}\ \emph {et~al.}(2014)\citenamefont {Aceti}, \citenamefont {Bayar}, \citenamefont {Oset}, \citenamefont {Martinez~Torres}, \citenamefont {Khemchandani}, \citenamefont {Dias}, \citenamefont {Navarra},\ and\ \citenamefont {Nielsen}}]{Aceti:2014uea}%
  \BibitemOpen
  \bibfield  {author} {\bibinfo {author} {\bibfnamefont {F.}~\bibnamefont {Aceti}}, \bibinfo {author} {\bibfnamefont {M.}~\bibnamefont {Bayar}}, \bibinfo {author} {\bibfnamefont {E.}~\bibnamefont {Oset}}, \bibinfo {author} {\bibfnamefont {A.}~\bibnamefont {Martinez~Torres}}, \bibinfo {author} {\bibfnamefont {K.~P.}\ \bibnamefont {Khemchandani}}, \bibinfo {author} {\bibfnamefont {J.~M.}\ \bibnamefont {Dias}}, \bibinfo {author} {\bibfnamefont {F.~S.}\ \bibnamefont {Navarra}},\ and\ \bibinfo {author} {\bibfnamefont {M.}~\bibnamefont {Nielsen}},\ }\href {https://doi.org/10.1103/PhysRevD.90.016003} {\bibfield  {journal} {\bibinfo  {journal} {Phys. Rev. D}\ }\textbf {\bibinfo {volume} {90}},\ \bibinfo {pages} {016003} (\bibinfo {year} {2014})},\ \Eprint {https://arxiv.org/abs/1401.8216} {arXiv:1401.8216 [hep-ph]} \BibitemShut {NoStop}%
\bibitem [{\citenamefont {Gamermann}\ \emph {et~al.}(2010)\citenamefont {Gamermann}, \citenamefont {Nieves}, \citenamefont {Oset},\ and\ \citenamefont {Ruiz~Arriola}}]{Gamermann:2009uq}%
  \BibitemOpen
  \bibfield  {author} {\bibinfo {author} {\bibfnamefont {D.}~\bibnamefont {Gamermann}}, \bibinfo {author} {\bibfnamefont {J.}~\bibnamefont {Nieves}}, \bibinfo {author} {\bibfnamefont {E.}~\bibnamefont {Oset}},\ and\ \bibinfo {author} {\bibfnamefont {E.}~\bibnamefont {Ruiz~Arriola}},\ }\href {https://doi.org/10.1103/PhysRevD.81.014029} {\bibfield  {journal} {\bibinfo  {journal} {Phys. Rev. D}\ }\textbf {\bibinfo {volume} {81}},\ \bibinfo {pages} {014029} (\bibinfo {year} {2010})},\ \Eprint {https://arxiv.org/abs/0911.4407} {arXiv:0911.4407 [hep-ph]} \BibitemShut {NoStop}%
\bibitem [{\citenamefont {Ling}\ \emph {et~al.}(2022)\citenamefont {Ling}, \citenamefont {Liu}, \citenamefont {Geng}, \citenamefont {Wang},\ and\ \citenamefont {Xie}}]{Ling:2021bir}%
  \BibitemOpen
  \bibfield  {author} {\bibinfo {author} {\bibfnamefont {X.-Z.}\ \bibnamefont {Ling}}, \bibinfo {author} {\bibfnamefont {M.-Z.}\ \bibnamefont {Liu}}, \bibinfo {author} {\bibfnamefont {L.-S.}\ \bibnamefont {Geng}}, \bibinfo {author} {\bibfnamefont {E.}~\bibnamefont {Wang}},\ and\ \bibinfo {author} {\bibfnamefont {J.-J.}\ \bibnamefont {Xie}},\ }\href {https://doi.org/10.1016/j.physletb.2022.136897} {\bibfield  {journal} {\bibinfo  {journal} {Phys. Lett. B}\ }\textbf {\bibinfo {volume} {826}},\ \bibinfo {pages} {136897} (\bibinfo {year} {2022})},\ \Eprint {https://arxiv.org/abs/2108.00947} {arXiv:2108.00947 [hep-ph]} \BibitemShut {NoStop}%
\bibitem [{\citenamefont {Du}\ \emph {et~al.}(2022)\citenamefont {Du}, \citenamefont {Albaladejo}, \citenamefont {Guo},\ and\ \citenamefont {Nieves}}]{Du:2022jjv}%
  \BibitemOpen
  \bibfield  {author} {\bibinfo {author} {\bibfnamefont {M.-L.}\ \bibnamefont {Du}}, \bibinfo {author} {\bibfnamefont {M.}~\bibnamefont {Albaladejo}}, \bibinfo {author} {\bibfnamefont {F.-K.}\ \bibnamefont {Guo}},\ and\ \bibinfo {author} {\bibfnamefont {J.}~\bibnamefont {Nieves}},\ }\href {https://doi.org/10.1103/PhysRevD.105.074018} {\bibfield  {journal} {\bibinfo  {journal} {Phys. Rev. D}\ }\textbf {\bibinfo {volume} {105}},\ \bibinfo {pages} {074018} (\bibinfo {year} {2022})},\ \Eprint {https://arxiv.org/abs/2201.08253} {arXiv:2201.08253 [hep-ph]} \BibitemShut {NoStop}%
\bibitem [{\citenamefont {Dai}\ \emph {et~al.}(2023)\citenamefont {Dai}, \citenamefont {Song},\ and\ \citenamefont {Oset}}]{Dai:2023kwv}%
  \BibitemOpen
  \bibfield  {author} {\bibinfo {author} {\bibfnamefont {L.~R.}\ \bibnamefont {Dai}}, \bibinfo {author} {\bibfnamefont {J.}~\bibnamefont {Song}},\ and\ \bibinfo {author} {\bibfnamefont {E.}~\bibnamefont {Oset}},\ }\href {https://doi.org/10.1016/j.physletb.2023.138200} {\bibfield  {journal} {\bibinfo  {journal} {Phys. Lett. B}\ }\textbf {\bibinfo {volume} {846}},\ \bibinfo {pages} {138200} (\bibinfo {year} {2023})},\ \Eprint {https://arxiv.org/abs/2306.01607} {arXiv:2306.01607 [hep-ph]} \BibitemShut {NoStop}%
\bibitem [{\citenamefont {Mart{\'\i}nez~Torres}\ \emph {et~al.}(2020)\citenamefont {Mart{\'\i}nez~Torres}, \citenamefont {Khemchandani}, \citenamefont {Roca},\ and\ \citenamefont {Oset}}]{MartinezTorres:2020hus}%
  \BibitemOpen
  \bibfield  {author} {\bibinfo {author} {\bibfnamefont {A.}~\bibnamefont {Mart{\'\i}nez~Torres}}, \bibinfo {author} {\bibfnamefont {K.~P.}\ \bibnamefont {Khemchandani}}, \bibinfo {author} {\bibfnamefont {L.}~\bibnamefont {Roca}},\ and\ \bibinfo {author} {\bibfnamefont {E.}~\bibnamefont {Oset}},\ }\href {https://doi.org/10.1007/s00601-020-01568-y} {\bibfield  {journal} {\bibinfo  {journal} {Few Body Syst.}\ }\textbf {\bibinfo {volume} {61}},\ \bibinfo {pages} {35} (\bibinfo {year} {2020})},\ \Eprint {https://arxiv.org/abs/2005.14357} {arXiv:2005.14357 [nucl-th]} \BibitemShut {NoStop}%
\bibitem [{\citenamefont {Liu}\ \emph {et~al.}(2024)\citenamefont {Liu}, \citenamefont {Pan}, \citenamefont {Liu}, \citenamefont {Wu}, \citenamefont {Lu},\ and\ \citenamefont {Geng}}]{Liu:2024uxn}%
  \BibitemOpen
  \bibfield  {author} {\bibinfo {author} {\bibfnamefont {M.-Z.}\ \bibnamefont {Liu}}, \bibinfo {author} {\bibfnamefont {Y.-W.}\ \bibnamefont {Pan}}, \bibinfo {author} {\bibfnamefont {Z.-W.}\ \bibnamefont {Liu}}, \bibinfo {author} {\bibfnamefont {T.-W.}\ \bibnamefont {Wu}}, \bibinfo {author} {\bibfnamefont {J.-X.}\ \bibnamefont {Lu}},\ and\ \bibinfo {author} {\bibfnamefont {L.-S.}\ \bibnamefont {Geng}},\ }\href@noop {} {\bibfield  {journal} {\bibinfo  {journal} {arXiv:2404.06399 [hep-ph]}\ } (\bibinfo {year} {2024})},\ \Eprint {https://arxiv.org/abs/2404.06399} {arXiv:2404.06399 [hep-ph]} \BibitemShut {NoStop}%
\bibitem [{\citenamefont {Ma}\ \emph {et~al.}(2019)\citenamefont {Ma}, \citenamefont {Wang},\ and\ \citenamefont {Mei{\ss}ner}}]{Ma:2017ery}%
  \BibitemOpen
  \bibfield  {author} {\bibinfo {author} {\bibfnamefont {L.}~\bibnamefont {Ma}}, \bibinfo {author} {\bibfnamefont {Q.}~\bibnamefont {Wang}},\ and\ \bibinfo {author} {\bibfnamefont {U.-G.}\ \bibnamefont {Mei{\ss}ner}},\ }\href {https://doi.org/10.1088/1674-1137/43/1/014102} {\bibfield  {journal} {\bibinfo  {journal} {Chin. Phys. C}\ }\textbf {\bibinfo {volume} {43}},\ \bibinfo {pages} {014102} (\bibinfo {year} {2019})},\ \Eprint {https://arxiv.org/abs/1711.06143} {arXiv:1711.06143 [hep-ph]} \BibitemShut {NoStop}%
\bibitem [{\citenamefont {Ren}\ \emph {et~al.}(2018)\citenamefont {Ren}, \citenamefont {Malabarba}, \citenamefont {Geng}, \citenamefont {Khemchandani},\ and\ \citenamefont {Mart\'\i{}nez~Torres}}]{Ren:2018pcd}%
  \BibitemOpen
  \bibfield  {author} {\bibinfo {author} {\bibfnamefont {X.-L.}\ \bibnamefont {Ren}}, \bibinfo {author} {\bibfnamefont {B.~B.}\ \bibnamefont {Malabarba}}, \bibinfo {author} {\bibfnamefont {L.-S.}\ \bibnamefont {Geng}}, \bibinfo {author} {\bibfnamefont {K.~P.}\ \bibnamefont {Khemchandani}},\ and\ \bibinfo {author} {\bibfnamefont {A.}~\bibnamefont {Mart\'\i{}nez~Torres}},\ }\href {https://doi.org/10.1016/j.physletb.2018.08.034} {\bibfield  {journal} {\bibinfo  {journal} {Phys. Lett. B}\ }\textbf {\bibinfo {volume} {785}},\ \bibinfo {pages} {112} (\bibinfo {year} {2018})},\ \Eprint {https://arxiv.org/abs/1805.08330} {arXiv:1805.08330 [hep-ph]} \BibitemShut {NoStop}%
\bibitem [{\citenamefont {Wu}\ \emph {et~al.}(2021)\citenamefont {Wu}, \citenamefont {Liu},\ and\ \citenamefont {Geng}}]{Wu:2020job}%
  \BibitemOpen
  \bibfield  {author} {\bibinfo {author} {\bibfnamefont {T.-W.}\ \bibnamefont {Wu}}, \bibinfo {author} {\bibfnamefont {M.-Z.}\ \bibnamefont {Liu}},\ and\ \bibinfo {author} {\bibfnamefont {L.-S.}\ \bibnamefont {Geng}},\ }\href {https://doi.org/10.1103/PhysRevD.103.L031501} {\bibfield  {journal} {\bibinfo  {journal} {Phys. Rev. D}\ }\textbf {\bibinfo {volume} {103}},\ \bibinfo {pages} {L031501} (\bibinfo {year} {2021})},\ \Eprint {https://arxiv.org/abs/2012.01134} {arXiv:2012.01134 [hep-ph]} \BibitemShut {NoStop}%
\bibitem [{\citenamefont {Wei}\ \emph {et~al.}(2022)\citenamefont {Wei}, \citenamefont {Shen},\ and\ \citenamefont {Xie}}]{Wei:2022jgc}%
  \BibitemOpen
  \bibfield  {author} {\bibinfo {author} {\bibfnamefont {X.}~\bibnamefont {Wei}}, \bibinfo {author} {\bibfnamefont {Q.-H.}\ \bibnamefont {Shen}},\ and\ \bibinfo {author} {\bibfnamefont {J.-J.}\ \bibnamefont {Xie}},\ }\href {https://doi.org/10.1140/epjc/s10052-022-10675-5} {\bibfield  {journal} {\bibinfo  {journal} {Eur. Phys. J. C}\ }\textbf {\bibinfo {volume} {82}},\ \bibinfo {pages} {718} (\bibinfo {year} {2022})},\ \Eprint {https://arxiv.org/abs/2205.12526} {arXiv:2205.12526 [hep-ph]} \BibitemShut {NoStop}%
\bibitem [{\citenamefont {Sanchez~Sanchez}\ \emph {et~al.}(2018)\citenamefont {Sanchez~Sanchez}, \citenamefont {Geng}, \citenamefont {Lu}, \citenamefont {Hyodo},\ and\ \citenamefont {Valderrama}}]{SanchezSanchez:2017xtl}%
  \BibitemOpen
  \bibfield  {author} {\bibinfo {author} {\bibfnamefont {M.}~\bibnamefont {Sanchez~Sanchez}}, \bibinfo {author} {\bibfnamefont {L.-S.}\ \bibnamefont {Geng}}, \bibinfo {author} {\bibfnamefont {J.-X.}\ \bibnamefont {Lu}}, \bibinfo {author} {\bibfnamefont {T.}~\bibnamefont {Hyodo}},\ and\ \bibinfo {author} {\bibfnamefont {M.~P.}\ \bibnamefont {Valderrama}},\ }\href {https://doi.org/10.1103/PhysRevD.98.054001} {\bibfield  {journal} {\bibinfo  {journal} {Phys. Rev. D}\ }\textbf {\bibinfo {volume} {98}},\ \bibinfo {pages} {054001} (\bibinfo {year} {2018})},\ \Eprint {https://arxiv.org/abs/1707.03802} {arXiv:1707.03802 [hep-ph]} \BibitemShut {NoStop}%
\bibitem [{\citenamefont {Martinez~Torres}\ \emph {et~al.}(2019)\citenamefont {Martinez~Torres}, \citenamefont {Khemchandani},\ and\ \citenamefont {Geng}}]{MartinezTorres:2018zbl}%
  \BibitemOpen
  \bibfield  {author} {\bibinfo {author} {\bibfnamefont {A.}~\bibnamefont {Martinez~Torres}}, \bibinfo {author} {\bibfnamefont {K.~P.}\ \bibnamefont {Khemchandani}},\ and\ \bibinfo {author} {\bibfnamefont {L.-S.}\ \bibnamefont {Geng}},\ }\href {https://doi.org/10.1103/PhysRevD.99.076017} {\bibfield  {journal} {\bibinfo  {journal} {Phys. Rev. D}\ }\textbf {\bibinfo {volume} {99}},\ \bibinfo {pages} {076017} (\bibinfo {year} {2019})},\ \Eprint {https://arxiv.org/abs/1809.01059} {arXiv:1809.01059 [hep-ph]} \BibitemShut {NoStop}%
\bibitem [{\citenamefont {Wu}\ \emph {et~al.}(2019)\citenamefont {Wu}, \citenamefont {Liu}, \citenamefont {Geng}, \citenamefont {Hiyama},\ and\ \citenamefont {Valderrama}}]{Wu:2019vsy}%
  \BibitemOpen
  \bibfield  {author} {\bibinfo {author} {\bibfnamefont {T.-W.}\ \bibnamefont {Wu}}, \bibinfo {author} {\bibfnamefont {M.-Z.}\ \bibnamefont {Liu}}, \bibinfo {author} {\bibfnamefont {L.-S.}\ \bibnamefont {Geng}}, \bibinfo {author} {\bibfnamefont {E.}~\bibnamefont {Hiyama}},\ and\ \bibinfo {author} {\bibfnamefont {M.~P.}\ \bibnamefont {Valderrama}},\ }\href {https://doi.org/10.1103/PhysRevD.100.034029} {\bibfield  {journal} {\bibinfo  {journal} {Phys. Rev. D}\ }\textbf {\bibinfo {volume} {100}},\ \bibinfo {pages} {034029} (\bibinfo {year} {2019})},\ \Eprint {https://arxiv.org/abs/1906.11995} {arXiv:1906.11995 [hep-ph]} \BibitemShut {NoStop}%
\bibitem [{\citenamefont {Pang}\ \emph {et~al.}(2020)\citenamefont {Pang}, \citenamefont {Wu},\ and\ \citenamefont {Geng}}]{Pang:2020pkl}%
  \BibitemOpen
  \bibfield  {author} {\bibinfo {author} {\bibfnamefont {J.-Y.}\ \bibnamefont {Pang}}, \bibinfo {author} {\bibfnamefont {J.-J.}\ \bibnamefont {Wu}},\ and\ \bibinfo {author} {\bibfnamefont {L.-S.}\ \bibnamefont {Geng}},\ }\href {https://doi.org/10.1103/PhysRevD.102.114515} {\bibfield  {journal} {\bibinfo  {journal} {Phys. Rev. D}\ }\textbf {\bibinfo {volume} {102}},\ \bibinfo {pages} {114515} (\bibinfo {year} {2020})},\ \Eprint {https://arxiv.org/abs/2008.13014} {arXiv:2008.13014 [hep-lat]} \BibitemShut {NoStop}%
\bibitem [{\citenamefont {Xiao}\ \emph {et~al.}(2024)\citenamefont {Xiao}, \citenamefont {Pang},\ and\ \citenamefont {Wu}}]{Xiao:2024dyw}%
  \BibitemOpen
  \bibfield  {author} {\bibinfo {author} {\bibfnamefont {Q.-C.}\ \bibnamefont {Xiao}}, \bibinfo {author} {\bibfnamefont {J.-Y.}\ \bibnamefont {Pang}},\ and\ \bibinfo {author} {\bibfnamefont {J.-J.}\ \bibnamefont {Wu}},\ }\href@noop {} {\bibfield  {journal} {\bibinfo  {journal} {arXiv:2408.16590 [hep-lat]}\ } (\bibinfo {year} {2024})},\ \Eprint {https://arxiv.org/abs/2408.16590} {arXiv:2408.16590 [hep-lat]} \BibitemShut {NoStop}%
\bibitem [{\citenamefont {Li}\ \emph {et~al.}(2020)\citenamefont {Li} \emph {et~al.}}]{Belle:2020xca}%
  \BibitemOpen
  \bibfield  {author} {\bibinfo {author} {\bibfnamefont {Y.}~\bibnamefont {Li}} \emph {et~al.} (\bibinfo {collaboration} {Belle}),\ }\href {https://doi.org/10.1103/PhysRevD.102.112001} {\bibfield  {journal} {\bibinfo  {journal} {Phys. Rev. D}\ }\textbf {\bibinfo {volume} {102}},\ \bibinfo {pages} {112001} (\bibinfo {year} {2020})},\ \Eprint {https://arxiv.org/abs/2008.13341} {arXiv:2008.13341 [hep-ex]} \BibitemShut {NoStop}%
\bibitem [{\citenamefont {Tan}\ \emph {et~al.}(2024)\citenamefont {Tan}, \citenamefont {Liu}, \citenamefont {Chen}, \citenamefont {Yang}, \citenamefont {Huang},\ and\ \citenamefont {Ping}}]{Tan:2024omp}%
  \BibitemOpen
  \bibfield  {author} {\bibinfo {author} {\bibfnamefont {Y.}~\bibnamefont {Tan}}, \bibinfo {author} {\bibfnamefont {X.}~\bibnamefont {Liu}}, \bibinfo {author} {\bibfnamefont {X.}~\bibnamefont {Chen}}, \bibinfo {author} {\bibfnamefont {Y.}~\bibnamefont {Yang}}, \bibinfo {author} {\bibfnamefont {H.}~\bibnamefont {Huang}},\ and\ \bibinfo {author} {\bibfnamefont {J.}~\bibnamefont {Ping}},\ }\href {https://doi.org/10.1103/PhysRevD.110.016005} {\bibfield  {journal} {\bibinfo  {journal} {Phys. Rev. D}\ }\textbf {\bibinfo {volume} {110}},\ \bibinfo {pages} {016005} (\bibinfo {year} {2024})},\ \Eprint {https://arxiv.org/abs/2404.02048} {arXiv:2404.02048 [hep-ph]} \BibitemShut {NoStop}%
\bibitem [{\citenamefont {Zhang}\ \emph {et~al.}(2024)\citenamefont {Zhang}, \citenamefont {Hu}, \citenamefont {He}, \citenamefont {Liu}, \citenamefont {Shi}, \citenamefont {Lu},\ and\ \citenamefont {Wang}}]{Zhang:2024yfj}%
  \BibitemOpen
  \bibfield  {author} {\bibinfo {author} {\bibfnamefont {Z.}~\bibnamefont {Zhang}}, \bibinfo {author} {\bibfnamefont {X.-Y.}\ \bibnamefont {Hu}}, \bibinfo {author} {\bibfnamefont {G.}~\bibnamefont {He}}, \bibinfo {author} {\bibfnamefont {J.}~\bibnamefont {Liu}}, \bibinfo {author} {\bibfnamefont {J.-A.}\ \bibnamefont {Shi}}, \bibinfo {author} {\bibfnamefont {B.-N.}\ \bibnamefont {Lu}},\ and\ \bibinfo {author} {\bibfnamefont {Q.}~\bibnamefont {Wang}},\ }\href@noop {} {\bibfield  {journal} {\bibinfo  {journal} {arXiv:2409.01325 [hep-ph]}\ } (\bibinfo {year} {2024})},\ \Eprint {https://arxiv.org/abs/2409.01325} {arXiv:2409.01325 [hep-ph]} \BibitemShut {NoStop}%
\bibitem [{\citenamefont {Guo}\ \emph {et~al.}(2006)\citenamefont {Guo}, \citenamefont {Shen}, \citenamefont {Chiang}, \citenamefont {Ping},\ and\ \citenamefont {Zou}}]{Guo:2006fu}%
  \BibitemOpen
  \bibfield  {author} {\bibinfo {author} {\bibfnamefont {F.-K.}\ \bibnamefont {Guo}}, \bibinfo {author} {\bibfnamefont {P.-N.}\ \bibnamefont {Shen}}, \bibinfo {author} {\bibfnamefont {H.-C.}\ \bibnamefont {Chiang}}, \bibinfo {author} {\bibfnamefont {R.-G.}\ \bibnamefont {Ping}},\ and\ \bibinfo {author} {\bibfnamefont {B.-S.}\ \bibnamefont {Zou}},\ }\href {https://doi.org/10.1016/j.physletb.2006.08.064} {\bibfield  {journal} {\bibinfo  {journal} {Phys. Lett. B}\ }\textbf {\bibinfo {volume} {641}},\ \bibinfo {pages} {278} (\bibinfo {year} {2006})},\ \Eprint {https://arxiv.org/abs/hep-ph/0603072} {arXiv:hep-ph/0603072} \BibitemShut {NoStop}%
\bibitem [{\citenamefont {Gamermann}\ \emph {et~al.}(2007)\citenamefont {Gamermann}, \citenamefont {Oset}, \citenamefont {Strottman},\ and\ \citenamefont {Vicente~Vacas}}]{Gamermann:2006nm}%
  \BibitemOpen
  \bibfield  {author} {\bibinfo {author} {\bibfnamefont {D.}~\bibnamefont {Gamermann}}, \bibinfo {author} {\bibfnamefont {E.}~\bibnamefont {Oset}}, \bibinfo {author} {\bibfnamefont {D.}~\bibnamefont {Strottman}},\ and\ \bibinfo {author} {\bibfnamefont {M.~J.}\ \bibnamefont {Vicente~Vacas}},\ }\href {https://doi.org/10.1103/PhysRevD.76.074016} {\bibfield  {journal} {\bibinfo  {journal} {Phys. Rev. D}\ }\textbf {\bibinfo {volume} {76}},\ \bibinfo {pages} {074016} (\bibinfo {year} {2007})},\ \Eprint {https://arxiv.org/abs/hep-ph/0612179} {arXiv:hep-ph/0612179} \BibitemShut {NoStop}%
\bibitem [{\citenamefont {Valderrama}(2018)}]{Valderrama:2018knt}%
  \BibitemOpen
  \bibfield  {author} {\bibinfo {author} {\bibfnamefont {M.~P.}\ \bibnamefont {Valderrama}},\ }\href {https://doi.org/10.1103/PhysRevD.98.014022} {\bibfield  {journal} {\bibinfo  {journal} {Phys. Rev. D}\ }\textbf {\bibinfo {volume} {98}},\ \bibinfo {pages} {014022} (\bibinfo {year} {2018})},\ \Eprint {https://arxiv.org/abs/1805.05100} {arXiv:1805.05100 [hep-ph]} \BibitemShut {NoStop}%
\bibitem [{\citenamefont {Ren}\ and\ \citenamefont {Sun}(2019)}]{Ren:2018qhr}%
  \BibitemOpen
  \bibfield  {author} {\bibinfo {author} {\bibfnamefont {X.-L.}\ \bibnamefont {Ren}}\ and\ \bibinfo {author} {\bibfnamefont {Z.-F.}\ \bibnamefont {Sun}},\ }\href {https://doi.org/10.1103/PhysRevD.99.094041} {\bibfield  {journal} {\bibinfo  {journal} {Phys. Rev. D}\ }\textbf {\bibinfo {volume} {99}},\ \bibinfo {pages} {094041} (\bibinfo {year} {2019})},\ \Eprint {https://arxiv.org/abs/1812.09931} {arXiv:1812.09931 [hep-ph]} \BibitemShut {NoStop}%
\bibitem [{\citenamefont {Ikeno}\ \emph {et~al.}(2023)\citenamefont {Ikeno}, \citenamefont {Bayar},\ and\ \citenamefont {Oset}}]{Ikeno:2022jbb}%
  \BibitemOpen
  \bibfield  {author} {\bibinfo {author} {\bibfnamefont {N.}~\bibnamefont {Ikeno}}, \bibinfo {author} {\bibfnamefont {M.}~\bibnamefont {Bayar}},\ and\ \bibinfo {author} {\bibfnamefont {E.}~\bibnamefont {Oset}},\ }\href {https://doi.org/10.1103/PhysRevD.107.034006} {\bibfield  {journal} {\bibinfo  {journal} {Phys. Rev. D}\ }\textbf {\bibinfo {volume} {107}},\ \bibinfo {pages} {034006} (\bibinfo {year} {2023})},\ \Eprint {https://arxiv.org/abs/2208.03698} {arXiv:2208.03698 [hep-ph]} \BibitemShut {NoStop}%
\bibitem [{\citenamefont {Bayar}\ \emph {et~al.}(2023)\citenamefont {Bayar}, \citenamefont {Ikeno},\ and\ \citenamefont {Roca}}]{Bayar:2023itf}%
  \BibitemOpen
  \bibfield  {author} {\bibinfo {author} {\bibfnamefont {M.}~\bibnamefont {Bayar}}, \bibinfo {author} {\bibfnamefont {N.}~\bibnamefont {Ikeno}},\ and\ \bibinfo {author} {\bibfnamefont {L.}~\bibnamefont {Roca}},\ }\href {https://doi.org/10.1103/PhysRevD.107.054042} {\bibfield  {journal} {\bibinfo  {journal} {Phys. Rev. D}\ }\textbf {\bibinfo {volume} {107}},\ \bibinfo {pages} {054042} (\bibinfo {year} {2023})},\ \Eprint {https://arxiv.org/abs/2301.07436} {arXiv:2301.07436 [hep-ph]} \BibitemShut {NoStop}%
\bibitem [{\citenamefont {Ren}\ \emph {et~al.}(2020)\citenamefont {Ren}, \citenamefont {Khemchandani},\ and\ \citenamefont {Martinez~Torres}}]{Ren:2019rts}%
  \BibitemOpen
  \bibfield  {author} {\bibinfo {author} {\bibfnamefont {X.-L.}\ \bibnamefont {Ren}}, \bibinfo {author} {\bibfnamefont {K.~P.}\ \bibnamefont {Khemchandani}},\ and\ \bibinfo {author} {\bibfnamefont {A.}~\bibnamefont {Martinez~Torres}},\ }\href {https://doi.org/10.1103/PhysRevD.102.016005} {\bibfield  {journal} {\bibinfo  {journal} {Phys. Rev. D}\ }\textbf {\bibinfo {volume} {102}},\ \bibinfo {pages} {016005} (\bibinfo {year} {2020})},\ \Eprint {https://arxiv.org/abs/1912.03369} {arXiv:1912.03369 [hep-ph]} \BibitemShut {NoStop}%
\bibitem [{\citenamefont {Deloff}(2000)}]{Deloff:1999gc}%
  \BibitemOpen
  \bibfield  {author} {\bibinfo {author} {\bibfnamefont {A.}~\bibnamefont {Deloff}},\ }\href {https://doi.org/10.1103/PhysRevC.61.024004} {\bibfield  {journal} {\bibinfo  {journal} {Phys. Rev. C}\ }\textbf {\bibinfo {volume} {61}},\ \bibinfo {pages} {024004} (\bibinfo {year} {2000})}\BibitemShut {NoStop}%
\bibitem [{\citenamefont {Kamalov}\ \emph {et~al.}(2001)\citenamefont {Kamalov}, \citenamefont {Oset},\ and\ \citenamefont {Ramos}}]{Kamalov:2000iy}%
  \BibitemOpen
  \bibfield  {author} {\bibinfo {author} {\bibfnamefont {S.~S.}\ \bibnamefont {Kamalov}}, \bibinfo {author} {\bibfnamefont {E.}~\bibnamefont {Oset}},\ and\ \bibinfo {author} {\bibfnamefont {A.}~\bibnamefont {Ramos}},\ }\href {https://doi.org/10.1016/S0375-9474(00)00709-0} {\bibfield  {journal} {\bibinfo  {journal} {Nucl. Phys. A}\ }\textbf {\bibinfo {volume} {690}},\ \bibinfo {pages} {494} (\bibinfo {year} {2001})},\ \Eprint {https://arxiv.org/abs/nucl-th/0010054} {arXiv:nucl-th/0010054} \BibitemShut {NoStop}%
\bibitem [{\citenamefont {Martinez~Torres}\ \emph {et~al.}(2011)\citenamefont {Martinez~Torres}, \citenamefont {Garzon}, \citenamefont {Oset},\ and\ \citenamefont {Dai}}]{MartinezTorres:2010ax}%
  \BibitemOpen
  \bibfield  {author} {\bibinfo {author} {\bibfnamefont {A.}~\bibnamefont {Martinez~Torres}}, \bibinfo {author} {\bibfnamefont {E.~J.}\ \bibnamefont {Garzon}}, \bibinfo {author} {\bibfnamefont {E.}~\bibnamefont {Oset}},\ and\ \bibinfo {author} {\bibfnamefont {L.~R.}\ \bibnamefont {Dai}},\ }\href {https://doi.org/10.1103/PhysRevD.83.116002} {\bibfield  {journal} {\bibinfo  {journal} {Phys. Rev. D}\ }\textbf {\bibinfo {volume} {83}},\ \bibinfo {pages} {116002} (\bibinfo {year} {2011})},\ \Eprint {https://arxiv.org/abs/1012.2708} {arXiv:1012.2708 [hep-ph]} \BibitemShut {NoStop}%
\bibitem [{\citenamefont {Malabarba}\ \emph {et~al.}(2022)\citenamefont {Malabarba}, \citenamefont {Khemchandani},\ and\ \citenamefont {Torres}}]{Malabarba:2021taj}%
  \BibitemOpen
  \bibfield  {author} {\bibinfo {author} {\bibfnamefont {B.~B.}\ \bibnamefont {Malabarba}}, \bibinfo {author} {\bibfnamefont {K.~P.}\ \bibnamefont {Khemchandani}},\ and\ \bibinfo {author} {\bibfnamefont {A.~M.}\ \bibnamefont {Torres}},\ }\href {https://doi.org/10.1140/epja/s10050-022-00681-2} {\bibfield  {journal} {\bibinfo  {journal} {Eur. Phys. J. A}\ }\textbf {\bibinfo {volume} {58}},\ \bibinfo {pages} {33} (\bibinfo {year} {2022})},\ \Eprint {https://arxiv.org/abs/2103.09978} {arXiv:2103.09978 [hep-ph]} \BibitemShut {NoStop}%
\bibitem [{\citenamefont {Feijoo}\ \emph {et~al.}(2021)\citenamefont {Feijoo}, \citenamefont {Liang},\ and\ \citenamefont {Oset}}]{Feijoo:2021ppq}%
  \BibitemOpen
  \bibfield  {author} {\bibinfo {author} {\bibfnamefont {A.}~\bibnamefont {Feijoo}}, \bibinfo {author} {\bibfnamefont {W.~H.}\ \bibnamefont {Liang}},\ and\ \bibinfo {author} {\bibfnamefont {E.}~\bibnamefont {Oset}},\ }\href {https://doi.org/10.1103/PhysRevD.104.114015} {\bibfield  {journal} {\bibinfo  {journal} {Phys. Rev. D}\ }\textbf {\bibinfo {volume} {104}},\ \bibinfo {pages} {114015} (\bibinfo {year} {2021})},\ \Eprint {https://arxiv.org/abs/2108.02730} {arXiv:2108.02730 [hep-ph]} \BibitemShut {NoStop}%
\bibitem [{\citenamefont {Dai}\ \emph {et~al.}(2022)\citenamefont {Dai}, \citenamefont {Molina},\ and\ \citenamefont {Oset}}]{Dai:2021vgf}%
  \BibitemOpen
  \bibfield  {author} {\bibinfo {author} {\bibfnamefont {L.~R.}\ \bibnamefont {Dai}}, \bibinfo {author} {\bibfnamefont {R.}~\bibnamefont {Molina}},\ and\ \bibinfo {author} {\bibfnamefont {E.}~\bibnamefont {Oset}},\ }\href {https://doi.org/10.1103/PhysRevD.105.016029} {\bibfield  {journal} {\bibinfo  {journal} {Phys. Rev. D}\ }\textbf {\bibinfo {volume} {105}},\ \bibinfo {pages} {016029} (\bibinfo {year} {2022})},\ \Eprint {https://arxiv.org/abs/2110.15270} {arXiv:2110.15270 [hep-ph]} \BibitemShut {NoStop}%
\bibitem [{\citenamefont {Faddeev}(1961)}]{Faddeev:1960su}%
  \BibitemOpen
  \bibfield  {author} {\bibinfo {author} {\bibfnamefont {L.~D.}\ \bibnamefont {Faddeev}},\ }\href@noop {} {\bibfield  {journal} {\bibinfo  {journal} {Sov. Phys. JETP}\ }\textbf {\bibinfo {volume} {12}},\ \bibinfo {pages} {1014} (\bibinfo {year} {1961})}\BibitemShut {NoStop}%
\bibitem [{\citenamefont {Martinez~Torres}\ \emph {et~al.}(2008{\natexlab{a}})\citenamefont {Martinez~Torres}, \citenamefont {Khemchandani},\ and\ \citenamefont {Oset}}]{MartinezTorres:2007sr}%
  \BibitemOpen
  \bibfield  {author} {\bibinfo {author} {\bibfnamefont {A.}~\bibnamefont {Martinez~Torres}}, \bibinfo {author} {\bibfnamefont {K.~P.}\ \bibnamefont {Khemchandani}},\ and\ \bibinfo {author} {\bibfnamefont {E.}~\bibnamefont {Oset}},\ }\href {https://doi.org/10.1103/PhysRevC.77.042203} {\bibfield  {journal} {\bibinfo  {journal} {Phys. Rev. C}\ }\textbf {\bibinfo {volume} {77}},\ \bibinfo {pages} {042203} (\bibinfo {year} {2008}{\natexlab{a}})},\ \Eprint {https://arxiv.org/abs/0706.2330} {arXiv:0706.2330 [nucl-th]} \BibitemShut {NoStop}%
\bibitem [{\citenamefont {Martinez~Torres}\ \emph {et~al.}(2008{\natexlab{b}})\citenamefont {Martinez~Torres}, \citenamefont {Khemchandani}, \citenamefont {Geng}, \citenamefont {Napsuciale},\ and\ \citenamefont {Oset}}]{MartinezTorres:2008gy}%
  \BibitemOpen
  \bibfield  {author} {\bibinfo {author} {\bibfnamefont {A.}~\bibnamefont {Martinez~Torres}}, \bibinfo {author} {\bibfnamefont {K.~P.}\ \bibnamefont {Khemchandani}}, \bibinfo {author} {\bibfnamefont {L.~S.}\ \bibnamefont {Geng}}, \bibinfo {author} {\bibfnamefont {M.}~\bibnamefont {Napsuciale}},\ and\ \bibinfo {author} {\bibfnamefont {E.}~\bibnamefont {Oset}},\ }\href {https://doi.org/10.1103/PhysRevD.78.074031} {\bibfield  {journal} {\bibinfo  {journal} {Phys. Rev. D}\ }\textbf {\bibinfo {volume} {78}},\ \bibinfo {pages} {074031} (\bibinfo {year} {2008}{\natexlab{b}})},\ \Eprint {https://arxiv.org/abs/0801.3635} {arXiv:0801.3635 [nucl-th]} \BibitemShut {NoStop}%
\bibitem [{\citenamefont {Khemchandani}\ \emph {et~al.}(2008)\citenamefont {Khemchandani}, \citenamefont {Martinez~Torres},\ and\ \citenamefont {Oset}}]{Khemchandani:2008rk}%
  \BibitemOpen
  \bibfield  {author} {\bibinfo {author} {\bibfnamefont {K.~P.}\ \bibnamefont {Khemchandani}}, \bibinfo {author} {\bibfnamefont {A.}~\bibnamefont {Martinez~Torres}},\ and\ \bibinfo {author} {\bibfnamefont {E.}~\bibnamefont {Oset}},\ }\href {https://doi.org/10.1140/epja/i2008-10625-3} {\bibfield  {journal} {\bibinfo  {journal} {Eur. Phys. J. A}\ }\textbf {\bibinfo {volume} {37}},\ \bibinfo {pages} {233} (\bibinfo {year} {2008})},\ \Eprint {https://arxiv.org/abs/0804.4670} {arXiv:0804.4670 [nucl-th]} \BibitemShut {NoStop}%
\bibitem [{\citenamefont {Chand}\ and\ \citenamefont {Dalitz}(1962)}]{Chand:1962ec}%
  \BibitemOpen
  \bibfield  {author} {\bibinfo {author} {\bibfnamefont {R.}~\bibnamefont {Chand}}\ and\ \bibinfo {author} {\bibfnamefont {R.~H.}\ \bibnamefont {Dalitz}},\ }\href {https://doi.org/10.1016/0003-4916(62)90113-6} {\bibfield  {journal} {\bibinfo  {journal} {Annals Phys.}\ }\textbf {\bibinfo {volume} {20}},\ \bibinfo {pages} {1} (\bibinfo {year} {1962})}\BibitemShut {NoStop}%
\bibitem [{\citenamefont {Barrett}\ and\ \citenamefont {Deloff}(1999)}]{Barrett:1999cw}%
  \BibitemOpen
  \bibfield  {author} {\bibinfo {author} {\bibfnamefont {R.~C.}\ \bibnamefont {Barrett}}\ and\ \bibinfo {author} {\bibfnamefont {A.}~\bibnamefont {Deloff}},\ }\href {https://doi.org/10.1103/PhysRevC.60.025201} {\bibfield  {journal} {\bibinfo  {journal} {Phys. Rev. C}\ }\textbf {\bibinfo {volume} {60}},\ \bibinfo {pages} {025201} (\bibinfo {year} {1999})}\BibitemShut {NoStop}%
\bibitem [{\citenamefont {Toker}\ \emph {et~al.}(1981)\citenamefont {Toker}, \citenamefont {Gal},\ and\ \citenamefont {Eisenberg}}]{Toker:1981zh}%
  \BibitemOpen
  \bibfield  {author} {\bibinfo {author} {\bibfnamefont {G.}~\bibnamefont {Toker}}, \bibinfo {author} {\bibfnamefont {A.}~\bibnamefont {Gal}},\ and\ \bibinfo {author} {\bibfnamefont {J.~M.}\ \bibnamefont {Eisenberg}},\ }\href {https://doi.org/10.1016/0375-9474(81)90502-9} {\bibfield  {journal} {\bibinfo  {journal} {Nucl. Phys. A}\ }\textbf {\bibinfo {volume} {362}},\ \bibinfo {pages} {405} (\bibinfo {year} {1981})}\BibitemShut {NoStop}%
\bibitem [{\citenamefont {Gal}(2007)}]{Gal:2006cw}%
  \BibitemOpen
  \bibfield  {author} {\bibinfo {author} {\bibfnamefont {A.}~\bibnamefont {Gal}},\ }\href {https://doi.org/10.1142/S0217751X07035379} {\bibfield  {journal} {\bibinfo  {journal} {Int. J. Mod. Phys. A}\ }\textbf {\bibinfo {volume} {22}},\ \bibinfo {pages} {226} (\bibinfo {year} {2007})},\ \Eprint {https://arxiv.org/abs/nucl-th/0607067} {arXiv:nucl-th/0607067} \BibitemShut {NoStop}%
\bibitem [{\citenamefont {Roca}\ and\ \citenamefont {Oset}(2010)}]{Roca:2010tf}%
  \BibitemOpen
  \bibfield  {author} {\bibinfo {author} {\bibfnamefont {L.}~\bibnamefont {Roca}}\ and\ \bibinfo {author} {\bibfnamefont {E.}~\bibnamefont {Oset}},\ }\href {https://doi.org/10.1103/PhysRevD.82.054013} {\bibfield  {journal} {\bibinfo  {journal} {Phys. Rev. D}\ }\textbf {\bibinfo {volume} {82}},\ \bibinfo {pages} {054013} (\bibinfo {year} {2010})},\ \Eprint {https://arxiv.org/abs/1005.0283} {arXiv:1005.0283 [hep-ph]} \BibitemShut {NoStop}%
\bibitem [{\citenamefont {{Yamagata-Sekihara}}\ \emph {et~al.}(2010)\citenamefont {{Yamagata-Sekihara}}, \citenamefont {Roca},\ and\ \citenamefont {Oset}}]{Yamagata-Sekihara:2010muv}%
  \BibitemOpen
  \bibfield  {author} {\bibinfo {author} {\bibfnamefont {J.}~\bibnamefont {{Yamagata-Sekihara}}}, \bibinfo {author} {\bibfnamefont {L.}~\bibnamefont {Roca}},\ and\ \bibinfo {author} {\bibfnamefont {E.}~\bibnamefont {Oset}},\ }\href {https://doi.org/10.1103/PhysRevD.82.094017} {\bibfield  {journal} {\bibinfo  {journal} {Phys. Rev. D}\ }\textbf {\bibinfo {volume} {82}},\ \bibinfo {pages} {094017} (\bibinfo {year} {2010})},\ \Eprint {https://arxiv.org/abs/1010.0525} {arXiv:1010.0525 [hep-ph]} \BibitemShut {NoStop}%
\bibitem [{\citenamefont {Guo}\ \emph {et~al.}(2007)\citenamefont {Guo}, \citenamefont {Shen},\ and\ \citenamefont {Chiang}}]{Guo:2006rp}%
  \BibitemOpen
  \bibfield  {author} {\bibinfo {author} {\bibfnamefont {F.-K.}\ \bibnamefont {Guo}}, \bibinfo {author} {\bibfnamefont {P.-N.}\ \bibnamefont {Shen}},\ and\ \bibinfo {author} {\bibfnamefont {H.-C.}\ \bibnamefont {Chiang}},\ }\href {https://doi.org/10.1016/j.physletb.2007.01.050} {\bibfield  {journal} {\bibinfo  {journal} {Phys. Lett. B}\ }\textbf {\bibinfo {volume} {647}},\ \bibinfo {pages} {133} (\bibinfo {year} {2007})},\ \Eprint {https://arxiv.org/abs/hep-ph/0610008} {arXiv:hep-ph/0610008} \BibitemShut {NoStop}%
\bibitem [{\citenamefont {Mart\'\i{}nez~Torres}\ \emph {et~al.}(2015)\citenamefont {Mart\'\i{}nez~Torres}, \citenamefont {Oset}, \citenamefont {Prelovsek},\ and\ \citenamefont {Ramos}}]{MartinezTorres:2014kpc}%
  \BibitemOpen
  \bibfield  {author} {\bibinfo {author} {\bibfnamefont {A.}~\bibnamefont {Mart\'\i{}nez~Torres}}, \bibinfo {author} {\bibfnamefont {E.}~\bibnamefont {Oset}}, \bibinfo {author} {\bibfnamefont {S.}~\bibnamefont {Prelovsek}},\ and\ \bibinfo {author} {\bibfnamefont {A.}~\bibnamefont {Ramos}},\ }\href {https://doi.org/10.1007/JHEP05(2015)153} {\bibfield  {journal} {\bibinfo  {journal} {JHEP}\ }\textbf {\bibinfo {volume} {05}},\ \bibinfo {pages} {153}},\ \Eprint {https://arxiv.org/abs/1412.1706} {arXiv:1412.1706 [hep-lat]} \BibitemShut {NoStop}%
\bibitem [{\citenamefont {{Yamagata-Sekihara}}\ \emph {et~al.}(2011)\citenamefont {{Yamagata-Sekihara}}, \citenamefont {Nieves},\ and\ \citenamefont {Oset}}]{Yamagata-Sekihara:2010kpd}%
  \BibitemOpen
  \bibfield  {author} {\bibinfo {author} {\bibfnamefont {J.}~\bibnamefont {{Yamagata-Sekihara}}}, \bibinfo {author} {\bibfnamefont {J.}~\bibnamefont {Nieves}},\ and\ \bibinfo {author} {\bibfnamefont {E.}~\bibnamefont {Oset}},\ }\href {https://doi.org/10.1103/PhysRevD.83.014003} {\bibfield  {journal} {\bibinfo  {journal} {Phys. Rev. D}\ }\textbf {\bibinfo {volume} {83}},\ \bibinfo {pages} {014003} (\bibinfo {year} {2011})},\ \Eprint {https://arxiv.org/abs/1007.3923} {arXiv:1007.3923 [hep-ph]} \BibitemShut {NoStop}%
\bibitem [{\citenamefont {Gao}\ \emph {et~al.}(2022)\citenamefont {Gao} \emph {et~al.}}]{Belle:2021kub}%
  \BibitemOpen
  \bibfield  {author} {\bibinfo {author} {\bibfnamefont {X.~Y.}\ \bibnamefont {Gao}} \emph {et~al.} (\bibinfo {collaboration} {Belle}),\ }\href {https://doi.org/10.1103/PhysRevD.105.032002} {\bibfield  {journal} {\bibinfo  {journal} {Phys. Rev. D}\ }\textbf {\bibinfo {volume} {105}},\ \bibinfo {pages} {032002} (\bibinfo {year} {2022})},\ \Eprint {https://arxiv.org/abs/2112.02497} {arXiv:2112.02497 [hep-ex]} \BibitemShut {NoStop}%
\bibitem [{\citenamefont {Meng}\ \emph {et~al.}(2024)\citenamefont {Meng}, \citenamefont {Ortiz-Pacheco}, \citenamefont {Baru}, \citenamefont {Epelbaum}, \citenamefont {Padmanath},\ and\ \citenamefont {Prelovsek}}]{Meng:2024kkp}%
  \BibitemOpen
  \bibfield  {author} {\bibinfo {author} {\bibfnamefont {L.}~\bibnamefont {Meng}}, \bibinfo {author} {\bibfnamefont {E.}~\bibnamefont {Ortiz-Pacheco}}, \bibinfo {author} {\bibfnamefont {V.}~\bibnamefont {Baru}}, \bibinfo {author} {\bibfnamefont {E.}~\bibnamefont {Epelbaum}}, \bibinfo {author} {\bibfnamefont {M.}~\bibnamefont {Padmanath}},\ and\ \bibinfo {author} {\bibfnamefont {S.}~\bibnamefont {Prelovsek}},\ }\href@noop {} {\  (\bibinfo {year} {2024})},\ \Eprint {https://arxiv.org/abs/2411.06266} {arXiv:2411.06266 [hep-lat]} \BibitemShut {NoStop}%
\bibitem [{\citenamefont {Junnarkar}\ and\ \citenamefont {Mathur}(2024)}]{Junnarkar:2024kwd}%
  \BibitemOpen
  \bibfield  {author} {\bibinfo {author} {\bibfnamefont {P.~M.}\ \bibnamefont {Junnarkar}}\ and\ \bibinfo {author} {\bibfnamefont {N.}~\bibnamefont {Mathur}},\ }\href@noop {} {\  (\bibinfo {year} {2024})},\ \Eprint {https://arxiv.org/abs/2410.08519} {arXiv:2410.08519 [hep-lat]} \BibitemShut {NoStop}%
\bibitem [{\citenamefont {Francis}(2025)}]{Francis:2024fwf}%
  \BibitemOpen
  \bibfield  {author} {\bibinfo {author} {\bibfnamefont {A.}~\bibnamefont {Francis}},\ }\href {https://doi.org/10.1016/j.ppnp.2024.104143} {\bibfield  {journal} {\bibinfo  {journal} {Prog. Part. Nucl. Phys.}\ }\textbf {\bibinfo {volume} {140}},\ \bibinfo {pages} {104143} (\bibinfo {year} {2025})}\BibitemShut {NoStop}%
\bibitem [{\citenamefont {Whyte}\ \emph {et~al.}(2024)\citenamefont {Whyte}, \citenamefont {Wilson},\ and\ \citenamefont {Thomas}}]{Whyte:2024ihh}%
  \BibitemOpen
  \bibfield  {author} {\bibinfo {author} {\bibfnamefont {T.}~\bibnamefont {Whyte}}, \bibinfo {author} {\bibfnamefont {D.~J.}\ \bibnamefont {Wilson}},\ and\ \bibinfo {author} {\bibfnamefont {C.~E.}\ \bibnamefont {Thomas}},\ }\href@noop {} {\  (\bibinfo {year} {2024})},\ \Eprint {https://arxiv.org/abs/2405.15741} {arXiv:2405.15741 [hep-lat]} \BibitemShut {NoStop}%
\bibitem [{\citenamefont {Radhakrishnan}\ \emph {et~al.}(2024)\citenamefont {Radhakrishnan}, \citenamefont {Padmanath},\ and\ \citenamefont {Mathur}}]{Radhakrishnan:2024ihu}%
  \BibitemOpen
  \bibfield  {author} {\bibinfo {author} {\bibfnamefont {A.}~\bibnamefont {Radhakrishnan}}, \bibinfo {author} {\bibfnamefont {M.}~\bibnamefont {Padmanath}},\ and\ \bibinfo {author} {\bibfnamefont {N.}~\bibnamefont {Mathur}},\ }\href {https://doi.org/10.1103/PhysRevD.110.034506} {\bibfield  {journal} {\bibinfo  {journal} {Phys. Rev. D}\ }\textbf {\bibinfo {volume} {110}},\ \bibinfo {pages} {034506} (\bibinfo {year} {2024})},\ \Eprint {https://arxiv.org/abs/2404.08109} {arXiv:2404.08109 [hep-lat]} \BibitemShut {NoStop}%
\bibitem [{\citenamefont {Collins}\ \emph {et~al.}(2024)\citenamefont {Collins}, \citenamefont {Nefediev}, \citenamefont {Padmanath},\ and\ \citenamefont {Prelovsek}}]{Collins:2024sfi}%
  \BibitemOpen
  \bibfield  {author} {\bibinfo {author} {\bibfnamefont {S.}~\bibnamefont {Collins}}, \bibinfo {author} {\bibfnamefont {A.}~\bibnamefont {Nefediev}}, \bibinfo {author} {\bibfnamefont {M.}~\bibnamefont {Padmanath}},\ and\ \bibinfo {author} {\bibfnamefont {S.}~\bibnamefont {Prelovsek}},\ }\href {https://doi.org/10.1103/PhysRevD.109.094509} {\bibfield  {journal} {\bibinfo  {journal} {Phys. Rev. D}\ }\textbf {\bibinfo {volume} {109}},\ \bibinfo {pages} {094509} (\bibinfo {year} {2024})},\ \Eprint {https://arxiv.org/abs/2402.14715} {arXiv:2402.14715 [hep-lat]} \BibitemShut {NoStop}%
\bibitem [{\citenamefont {Padmanath}\ \emph {et~al.}(2024)\citenamefont {Padmanath}, \citenamefont {Radhakrishnan},\ and\ \citenamefont {Mathur}}]{Padmanath:2023rdu}%
  \BibitemOpen
  \bibfield  {author} {\bibinfo {author} {\bibfnamefont {M.}~\bibnamefont {Padmanath}}, \bibinfo {author} {\bibfnamefont {A.}~\bibnamefont {Radhakrishnan}},\ and\ \bibinfo {author} {\bibfnamefont {N.}~\bibnamefont {Mathur}},\ }\href {https://doi.org/10.1103/PhysRevLett.132.201902} {\bibfield  {journal} {\bibinfo  {journal} {Phys. Rev. Lett.}\ }\textbf {\bibinfo {volume} {132}},\ \bibinfo {pages} {201902} (\bibinfo {year} {2024})},\ \Eprint {https://arxiv.org/abs/2307.14128} {arXiv:2307.14128 [hep-lat]} \BibitemShut {NoStop}%
\bibitem [{\citenamefont {Aoki}\ \emph {et~al.}(2023)\citenamefont {Aoki}, \citenamefont {Aoki},\ and\ \citenamefont {Inoue}}]{Aoki:2023nzp}%
  \BibitemOpen
  \bibfield  {author} {\bibinfo {author} {\bibfnamefont {T.}~\bibnamefont {Aoki}}, \bibinfo {author} {\bibfnamefont {S.}~\bibnamefont {Aoki}},\ and\ \bibinfo {author} {\bibfnamefont {T.}~\bibnamefont {Inoue}},\ }\href {https://doi.org/10.1103/PhysRevD.108.054502} {\bibfield  {journal} {\bibinfo  {journal} {Phys. Rev. D}\ }\textbf {\bibinfo {volume} {108}},\ \bibinfo {pages} {054502} (\bibinfo {year} {2023})},\ \Eprint {https://arxiv.org/abs/2306.03565} {arXiv:2306.03565 [hep-lat]} \BibitemShut {NoStop}%
\end{thebibliography}%

\end{document}